\documentclass[prd,
twocolumn,
superscriptaddress,
nofootinbib,
]{revtex4-2}

\usepackage{graphicx}
\usepackage{bm}
\usepackage{csquotes}
\usepackage{float}
\usepackage{amsmath}
\usepackage{bm}
\usepackage{amssymb}
\usepackage{url}
\usepackage{setspace}
\usepackage{latexsym}
\usepackage{gensymb}
\usepackage{soul}
\usepackage{footnote}
\usepackage{makecell}
\usepackage{booktabs}
\usepackage{orcidlink}
\usepackage{fontawesome}

\usepackage{aas_macros}

\usepackage{hyperref}

\usepackage[dvipsnames]{xcolor}

\newcommand{\citationneeded}[1][]{{\color{red} [citation!]}}

\usepackage{placeins}
\begin{document}

\title{Prospects for High-Frequency Gravitational-Wave Detection with GEO600}

\author{Christopher M. Jungkind\,\orcidlink{0009-0003-6270-7765}}
\email{chrisjungkind@gmail.com}
\affiliation{\href{https://ror.org/04ytb9n23}{Furman University}, Greenville, SC 29613, USA}
\affiliation{Department of Physics and Astronomy, \href{https://ror.org/00mkhxb43}{University of Notre Dame}, IN 46556, USA}

\author{Brian C. Seymour\,\orcidlink{0000-0002-7865-1052}}
\affiliation{TAPIR, Walter Burke Institute for Theoretical Physics, \href{https://ror.org/05dxps055}{California Institute of Technology}, Pasadena, CA 91125, USA}

\author{Andrew Laeuger\,\orcidlink{0000-0002-8212-6496}}
\affiliation{TAPIR, Walter Burke Institute for Theoretical Physics, \href{https://ror.org/05dxps055}{California Institute of Technology}, Pasadena, CA 91125, USA}

\author{Yanbei Chen\,\orcidlink{0000-0002-9730-9463}}
\affiliation{TAPIR, Walter Burke Institute for Theoretical Physics, \href{https://ror.org/05dxps055}{California Institute of Technology}, Pasadena, CA 91125, USA}

\date{\today}

\begin{abstract}
Current ground-based interferometers are optimized for sensitivity from a few tens of Hz to about 1 kHz.
While they are not currently utilized for GW detection, interferometric detectors also feature narrow bands of strong sensitivity at higher frequencies where the sideband fields created by a GW are resonantly amplified in the optical system.
Small changes to system parameters allow the narrow band of high sensitivity to be scanned over a much larger range of frequencies.
In this paper, we investigate whether simply modifying the detuning angle of the signal-recycling mirror of the GEO600 interferometer can make this experiment sensitive to GWs in the kilohertz frequency range.
We compute the strain sensitivity for GEO600 across a frequency range from several kHz to tens of kHz for various detuning angles.
We also show that LIGO cannot attain the same effect assuming that the optical components are not changed due to the narrow band response of the Fabry-Perot cavities.
We then calculate the sensitivity of GEO600 to various proposed high-frequency GW sources and compare it to the sensitivity of other ground-based detectors. \href{https://github.com/Chris19j/GEO600-High-Frequency-Modeling}{\faGithub}
\end{abstract}

\maketitle

\section{\label{sec:introduction}Introduction}

The era of gravitational-wave (GW) astronomy began in 2015 with the first detection of a binary black-hole (BBH) merger by the LIGO observatories and has flourished since then with more than 100 observed compact merger events \cite{LIGOScientific:2014pky, LIGOScientific:2016aoc, VIRGO:2014yos, LIGOScientific:2017vwq, KAGRA:2021vkt}. Despite these achievements, current GW detectors are only sensitive between a few tens of Hz and several kilohertz (kHz). The high frequency sensitivity is set primarily by quantum shot noise. While shot noise is generally frequency-independent (white noise), the response of the interferometer to GWs weakens with increasing frequency in the low kHz regime, resulting in poorer sensitivity \cite{Maggiore:2008ulw,PhysRevD.96.084004,Rakhmanov_2008}. Thus, the pathway to detecting high frequency GWs lies in either building new detectors or modifying the response of existing detectors.

Recent years have yielded a diverse set of proposals to improve the high frequency sensitivity of GW detectors. The LIGO and Virgo interferometers have developed both frequency-independent and frequency-dependent squeezing technologies, though in the O3, O4a and O4b observing runs, Virgo has only employed frequency-independent squeezing while LIGO has operated with frequency-dependent squeezing \cite{Tse:2019wcy, LIGOO4Detector:2023wmz,VIRGOsqueezing}.
LIGO A+, Voyager, and A$\#$ are design concepts that aim to retrofit the existing LIGO observatory facilities by implementing  upgraded technology \cite{Aplus, Adhikari_2020, KAGRA:2013rdx}. Third generation ground interferometers \cite{einsteinTelescope, Reitze:2019iox} will expand the sensitivity range of such experiments. At higher frequencies, resonant spheres \cite{sphereResonators, resonantSpheres} could detect GWs up to $10$ kHz, optically-levitated sensors \cite{PhysRevLett.128.111101, PhysRevLett.110.071105} will search for GWs at tens to hundreds of kHz, and resonant electromagnetic detectors \cite{Herman_2021, Herman_2023, Domcke_2022, radioGWdetector, SRFcavities}, bulk acoustic wave devices \cite{Goryachev_2014, Goryachev_2021, Page_2021}, and tabletop-scale interferometers \cite{Akutsu_2008, QUEST, decameterMI} will probe frequencies at MHz and above. Additionally, space-based interferometers \cite{LISA:2017pwj, TianGO, DECIGO}, atom interferometry experiments \cite{MAGIS-100, AGIS, atomInterferometer}, and pulsar timing arrays (PTAs) \cite{NANOgrav, PTAdata} will target frequencies below $1$ Hz. While such technologies for expanding the range of detectable GW frequencies are very promising --- indeed, PTAs have recently yielded strong evidence for the existence of a stochastic GW background \cite{NANOGrav:2023gor,Antoniadis:2022pcn} --- the vast majority will require at least a decade of development to become operational.

Detecting GWs at frequencies in the tens of kHz holds the potential to unveil signatures of new physics such as very light compact object binaries or ultralight boson clouds \cite{andrewsGeniusBreakoutPaper, Aggarwal:2020olq}.
Sub-solar mass (SSM) compact objects could be astrophysical objects like BHs created through an unconventional formation channel \cite{PhysRevLett.121.221102, MACHObinaries, darkhaloBlackHoles}, a primordial BH \cite{Zeldovich:1967lct, Hawking:1971ei, 1974MNRAS.168..399C}, or some other exotic compact object \cite{bosonStar, gravastars, gravitinoStars, moduliStars}.
If multiple SSM objects form a binary, it will be detectable by searching for its effects with a high-frequency GW detector \cite{Shandera:2018xkn, Miller:2024rca}. Additionally, isolated spinning black holes can form boson clouds via the superradiance effect if there is some unknown ultralight boson cloud \cite{Arvanitaki:2014wva,Brito:2017zvb}. Detection of the GWs emitted by the boson clouds could be evidence of new physics. The quantum chromodynamics (QCD) axion \cite{Peccei:1977hh,Peccei:1977ur,Wilczek:1977pj,Weinberg:1977ma} and the dark photon \cite{Goodsell:2009xc, Petraki:2013wwa,Fabbrichesi:2020wbt} are two prominent ultralight bosons that could have astrophysical effects in GWs \cite{PhysRevD.95.043541, coldDarkMatter, RevModPhys.90.045002}. The LVK collaboration has actively searched for GWs from SSM binaries \cite{LIGOScientific:2018glc, LIGOScientific:2019kan, LIGOScientific:2021job, Nitz:2021vqh, nancyPBH} and superradiant boson clouds in their data \cite{scalarBCLIGOsearch, Sun:2019mqb, darkPhoton}. 
SSM binaries and superradiant boson clouds may generate GWs with characteristic frequencies above the sensitivity bands of current generation detectors \cite{Franciolini:2022htd}.
In addition to superradiance, ultralight bosons could be bound to the gravitational potential of a NS \cite{Goldman:1989nd,Kouvaris:2010vv} or BH which would allow indirect detection via fifth-force in GW signals \cite{Hook:2017psm,Huang:2018pbu,Alexander:2018qzg} or in binary pulsar observations \cite{Stairs:2003eg,Wex:2014nva,Seymour:2019tir,Seymour:2020yle}.
Outside of GW observations, ultralight bosons and SSM compact objects have been searched for with both direct and indirect experiments; however, no evidence has been found \cite{PhysRevLett.104.041301, PhysRevLett.111.251301, ARMENGAUD2011329, Abdo_2010, Atwood_2009, collaboration_2015, EROS-2:2006ryy, Lasserre:2000xw}. Lastly, enhancing the high-frequency sensitivity of GW detectors expands the range of astrophysical phenomena that can be probed \cite{Dhani:2025xno}.

As a more near term solution to detecting these high-frequency sources, we consider modifications to existing GW detectors that could enhance their sensitivity in the $\sim5-100$ kHz range. When the GW period approaches the photon round-trip time in optical cavities within an interferometer, the sideband fields produced by a GW can be resonantly amplified inside optical cavities within the experiment \cite{Aggarwal:2025noe}.~This sideband amplification offers special frequencies at which the sensitivity of the detector could be greatly improved. In this work, we investigate whether GEO600 \cite{Dooley_2016}, a dual-recycled, folded-arm, Michelson interferometer (MI) \cite{Willke:2002bs, Hild:2007DC, Freise:2003xpa}, can be made sensitive to sources in the tens of kHz frequency range by utilizing these special frequencies.

The rest of this paper is organized as follows. In Sec.~\ref{sec:Detector-Sensitivity}, we discuss the underlying detector physics that motivated the investigation of GEO600 as a high-frequency GW detector. We describe our methods for computing strain sensitivities for detuned configurations of GEO600. We discuss the resulting sensitivities of different interferometer configurations and compare them to other ground-based interferometric detectors. In Sec.~\ref{sec:high-frequency-sources}, we introduce the high-frequency astrophysical sources and describe their GW waveforms and properties. We discuss how the signal-to-noise ratio (SNR) is calculated and the resulting ability of each interferometer to detect the sources. In Sec.~\ref{sec:conclusion}, we conclude our work and discuss future research. The data used in this paper can be downloaded from GitHub \cite{OurCode}. 

\section{\label{sec:Detector-Sensitivity}Detector Sensitivity}
In this section, we outline the principles behind narrow-band high-frequency GW detection via resonant enhancement of GW-induced sidebands. We then outline our procedure for computing the strain sensitivity of ground-based interferometric detectors given their optical configuration.

\subsection{\label{subsec:Underlying-Physics}Underlying Physics}
\begin{figure}
    \centering 
    \includegraphics[width=\columnwidth]{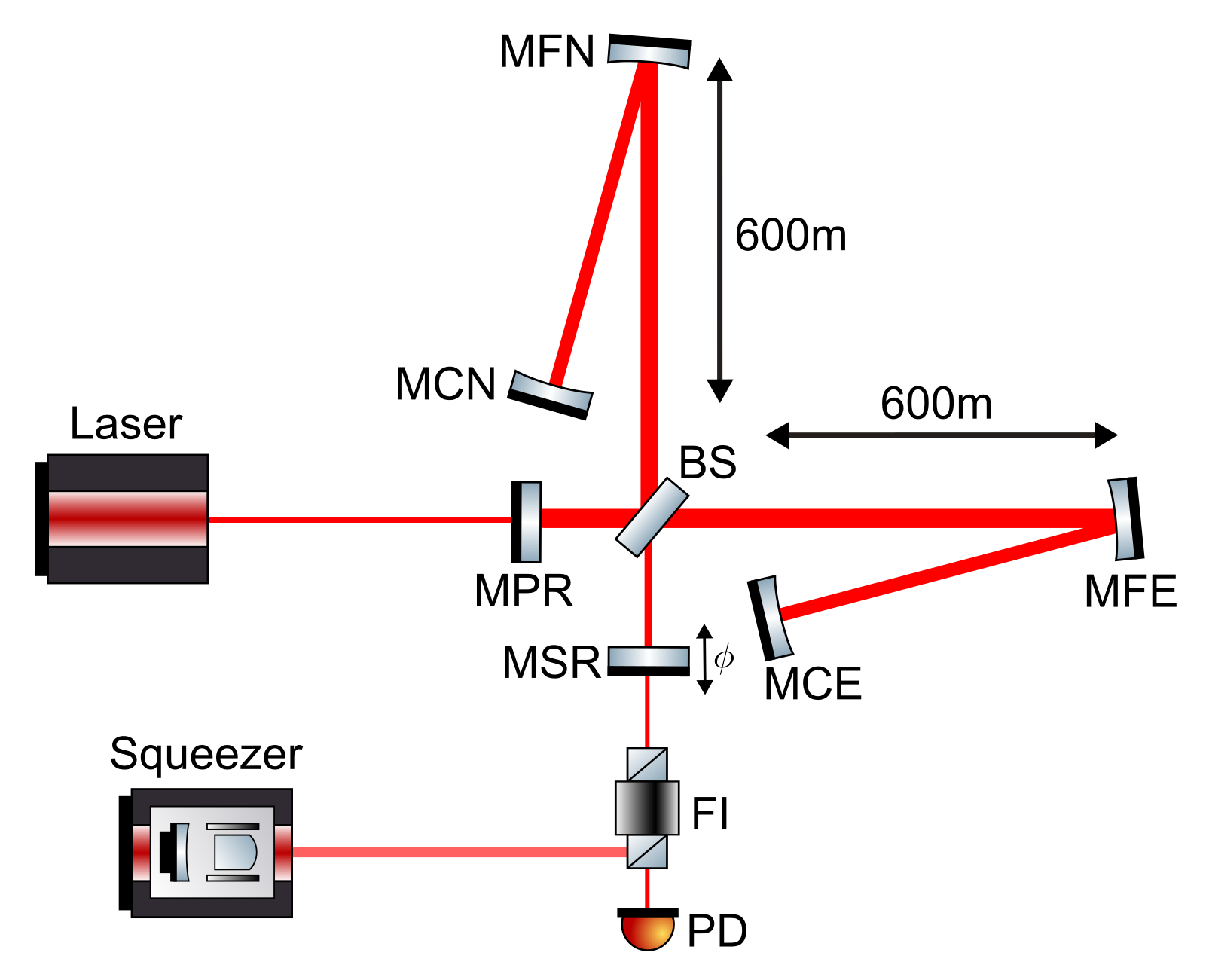} 
    \caption{GEO600 optical layout used in Finesse model. FI: Faraday isolator, MPR: power-recycling mirror, MSR: signal-recycling mirror, BS: beam splitter, PD: photodiode for DC readout. MFE and MCE are the far east mirror and central east mirror. $\phi$ is the detuning angle of the MSR in degrees of a wavelength. The notation for the north arm is analogous.}
    \label{GEO-optical-layout}
\end{figure}

To improve the sensitivity of a GW detector, one can either weaken the background noise sources or amplify the magnitude of the detector's response to an incident GW relative to the response to perturbations in other channels. GEO600's potential as a high-frequency GW detector originates in the latter option, namely through the resonant enhancement of GW-induced sideband fields within the interferometer \cite{Grote:2004sf}. Resonant amplification of the sideband fields in turn amplifies the signal power at the detection port, boosting the overall response. GEO600 features two mirrors which reflect laser light departing the Michelson arms back into the MI, thus forming effective cavities with the north and east Michelson arm mirrors (shown in Fig.~\ref{GEO-optical-layout}).~The power-recycling cavity (PRC), created by the power-recycling mirror (MPR) and MI, increases the sensitivity of the interferometer by building up the cavity light’s intensity and distance traveled \cite{FP_cavity}.~The signal-recycling cavity (SRC), created by the signal-recycling mirror (MSR) and MI, allows the signal sidebands to resonate at specific Fourier frequencies, denoted resonant frequencies \cite{originalDualRecycling, MIZUNO1993273,HEINZEL1996305}.

The resonant frequencies in the SRC are those which experience a net phase shift of $2\pi n$ during a round trip in this effective cavity. When the MSR is detuned by angle $\phi$, the SRC resonates for sideband frequencies $f$ which satisfy
\begin{equation}
    2\times (2\pi(\nu+f)L_\mathrm{SRC}/c+\phi)=2\pi n,
\end{equation}
where $\nu$ is the laser frequency, and $n$ is an integer. For GEO600, $L_{\mathrm{SRC}} = L_{\mathrm{MI}} + L_{\mathrm{SR}}$ is the optical distance light travels from the MI's end mirror to the MSR (when the detuning is zero). For this folded configuration, $L_{\mathrm{MI}}=2\times 600=1200$m. The definition of $L_{\mathrm{SRC}}$ ensures that $2\nu L_{\mathrm{SRC}}/c$ is an integer already; since GWs create sidebands at $\pm f_\mathrm{GW}$, the condition simplifies to one for the (positive) GW frequencies, which produce sidebands that are resonant in the SRC:
\begin{equation}
    f_\mathrm{res}=\frac{c}{2L_\mathrm{SRC}}\Bigr|n-\frac{\phi}{\pi}\Bigr|\, .
\end{equation}
For this work, we primarily target the frequency which appears for $n=0$\footnote{The resonant frequencies for $|n|>0$ are less advantageous for detection because the GW-induced phase delay in the MI is diminished for frequencies greater than $f_\mathrm{fsr} = c/2L_\mathrm{src}$ (see Appendix~\ref{Appendix-A}).}. The detuning angle $\phi$ is related to the MSR displacement $\delta l$ by $\phi=k\delta l$, where $k = 2\pi \nu/c$. Therefore, by finely detuning the MSR location, we can scan the SRC resonant frequencies, and thus the resonant peaks in the GEO600 detector response, over a broad frequency range \cite{Heinzel:1999rk}. Detuning the MSR allows for lowest-order resonant peaks in the detector response function to occur as high as 62.5 kHz.

Unlike GEO600, the LIGO detector response displays both Fabry-Perot and signal-recycling resonant features, but we will see that it is not well suited for high-frequency detection with current mirror properties. The Fabry-Perot resonant feature occurs at multiples of the FSR of the Fabry-Perot arms (37.5 kHz), but it is unaffected by the detuning of the MSR, and thus cannot be scanned with this procedure. One might envision detuning the Fabry-Perot mirrors themselves, but this would drastically reduce the DC power circulating in the arms (that is, intentionally moving the cavities away from their locked position), heavily attenuating the available power for sideband generation. 

The signal-recycling resonant feature can be scanned by detuning the SRC; however, it cannot reach into our target frequency band. From a optical resonance perspective, the Fabry-Perot arm cavities increase the effective SRC length by a factor of $\sim 2\mathcal{F}$, where $\mathcal{F}$ is the cavity finesse. This limits the maximum first-order resonance frequency to $\sim 200$ Hz with current LIGO mirror transmissivities; all higher-order resonant frequencies of the SRC are washed out by the high-finesse Fabry-Perot arms.

As it turns out, though, this feature is not produced exclusively by optical resonance behavior. In fact, in the small detuning angle regime, the frequency of optimal sensitivity (in the $\sim 100$ Hz region) scales as $\sqrt{\phi P_\mathrm{laser}}$, indicative of the presence of an optical spring effect \cite{BuonannoOpticalSpring}. The optical spring creates sharp features in the aLIGO sensitivity function for small detuning angles. However, as the detuning angle increases, the optical spring frequency increases, but its linewidth increases at a faster rate, resulting in a broader feature. The optical spring frequency is also limited to around 200 Hz with the current properties of LIGO optics, so neither the optical resonance nor optical spring effects can generate improvements to the LIGO sensitivity function outside its nominal frequency band simply by detuning the signal recycling mirror. Thus, a GEO600-like detector geometry --- namely, one featuring shorter arms (and thus a higher FSR) and a lack of Fabry-Perot cavities --- is more favorable for high-frequency detection via the procedure we have described.

For clarity, we define the interferometer terminology used throughout this paper. The \textit{tuned} interferometer configuration corresponds to a MSR detuning angle of $\phi=0$ for GEO600 and $\phi = 90$ for aLIGO. The \textit{anti-tuned} configuration corresponds to a MSR detuning angle of $\phi=90$ for GEO600 and $\phi=0$ for aLIGO\footnote{In practice for this configuration, we use $\phi=1$ since taking the limit as $\phi\rightarrow 0$ exactly corresponds to a configuration completely dominated by classical noise.}. A \textit{detuned} configuration corresponds to any MSR detuning angle between $0 < \phi < 90$ for GEO600 aLIGO. For GEO600, we use a \textit{scanned} configuration to show the pointwise optimization in frequency space over MSR detuning angles. This scanned configuration is used solely as a visualization tool, as it is not a physically utilizable sensitivity itself.

All the features of the detector response functions we have described can be seen in Figs. \ref{GEO-detuning-angles} and \ref{aLIGO-detuning-angles}, the generation of which is outlined in the following section.

\subsection{\label{subsec:Methods}Methods}

To construct the strain sensitivity function for a given configuration of GEO600, we include the following dominant noise sources $n_i$: laser quantum noise, classical laser amplitude and frequency noise, seismic noise, and thermal noise in the suspensions, substrates, and coatings of the optics. The expressions for the amplitude spectral densities for each noise source $A_{n_i}(f)$ (with units $[n_i]/\sqrt{\mathrm{Hz}}$) are presented in Appendix \ref{Appendix-B}. 

The total sky-averaged strain noise power spectral density (PSD) is given by
\begin{equation}
    \langle S_h(f) \rangle =\Bigr\langle\frac{\partial P}{\partial h}\Bigr\rangle^{-2}\sum_{n_i}S_{n_i}(f)
    \Bigr(\frac{\partial P}{\partial n_i}\Bigr)^2,
    \label{eq:strain-noise-PSD}
\end{equation}
where the summation occurs over all the noise sources $n_i$, $S_{n_i}\equiv A_{n_i}^2$ is the PSD of the noise source $n_i$, and $\partial P/\partial n_i$ is the transfer function from noise source $n_i$ to the detection observable $P$. In this paper, we designate the observable as the power at the dark port of the interferometer --- see the optical layout depicted in Fig.~\ref{GEO-optical-layout}. Lastly, $\langle\partial P/\partial h\rangle$ is the transfer function from an incident GW to the observable (i.e., the detector response function). We use the notation $\langle \cdot \rangle$ to represent a sky-averaged quantity. To summarize, the transfer function $\partial P/\partial n_i$ projects a given noise source onto the readout, and $\partial P/\partial h$ projects the GW strain onto the same readout; thus, Eq. \eqref{eq:strain-noise-PSD} projects each noise source into some equivalent sky-averaged strain noise.

\begin{table}
    \begin{tabular}{lccc}
         \toprule
         \toprule
         \addlinespace[5pt]
         Mirror & Transmission & Loss & Mass
         \\
         \toprule
         \makecell[l]{Beam splitter \\ (BS)} & $0.513872$  & $130$ ppm & $9.3$ kg
         \\
         \makecell[l]{Central East mirror \\(MCE)} & $13$ ppm & $130$ ppm & $5.6$ kg
         \\
         \makecell[l]{Central North mirror \\(MCN)} & $13$ ppm & $130$ ppm & $5.6$ kg
         \\
         \makecell[l]{Far East mirror \\(MFE)} & $8.3$ ppm & $130$ ppm & $5.6$ kg
         \\
         \makecell[l]{Far North mirror \\(MFN)} & $8.3$ ppm & $130$ ppm & $5.6$ kg
         \\
         \makecell[l]{Power-recycling mirror \\(MPR)} & $900$ ppm & $130$ ppm & $2.92$ kg
         \\
         \makecell[l]{Signal-recycling mirror \\(MSR)} & $0.09995$ & $50$ ppm & $2.92$ kg
         \\
         \toprule
         \toprule
    \end{tabular}
    \caption{\label{tab:table1} Mirror parameters used in the GEO600 Finesse model \cite{Wittel:2009xia, Gossler:2004bbb, Affeldt:2014ddd, LoughPrivate}.~Ppm: parts per million, kg: kilogram.}
\end{table}

To compute $S_h(f)$, we employ the Finesse 3.0 package \cite{Freise:2003ca,Freise:2009sf}, which we use to extract the frequency-domain transfer functions $\partial P/\partial n_i$\footnote{Finesse features a method to compute the quantum noise at any optical port, so for quantum noise in particular, we directly extract the noise at the detection port.} and $\partial P/\partial h$ from a model of GEO600. The numerical methods of Finesse have been validated and do not need low-frequency approximations \cite{Freise:2009sf}. The transfer functions from a GW to an observable are computed in Finesse under the assumption that the GW is incident from directly above the interferometer. Denoting the transfer function to the dark port power calculated by Finesse as $\frac{\partial P}{\partial h}\bigr |_{\texttt{Finesse}}$, the sky-averaged detector response function is therefore equal to

\begin{equation}
   \left\langle \frac{\partial P}{\partial h} \right\rangle = \frac{\partial P}{\partial h}\biggr |_{\texttt{Finesse}}\!\!\!\times \frac{\mathcal{R}(f)}{\left|F_+^\mathrm{fold}\left(f,\hat n = \hat z\right)\right|} \, ,
\end{equation}
where $\mathcal{R}(f)$ is the sky-averaged detector antenna pattern and $\left|F_+^\mathrm{fold}\left(f,\hat n = \hat z\right)\right|$ is the antenna pattern function evaluated for a GW incident from overhead. In the low-frequency limit $\mathcal{R}(f)=2/5$, but for frequencies comparable to the inverse light travel time in the Michelson arms, high frequency effects become important.
In Appendix \ref{Appendix-A}, we derive the antenna pattern of a folded Michelson interferometer as done in Ref.~\cite{LoughHFpaper} and clarify how the folded geometry modifies the high-frequency response. 

Note that while we use sky averaging for SNR/detectability calculations, when we plot detector strain sensitivities in this work (e.g., Figs. \ref{GEO-detuning-angles}, \ref{aLIGO-detuning-angles}, \ref{strain-sensitivity-plot}, and \ref{Fig.6}), we assume a fiducial GW sky location of $\theta_{\hat n} =15$ and $\phi_{\hat n}=0$, as is typically done in the CE community \cite{Srivastava:2022slt,Evans:2021gyd}.

Table~\ref{tab:table1} displays the parameters of the important mirrors in GEO600 needed to model the experiment accurately. We model the detection readout method using a DC readout \cite{Affeldt:2014hjx} and maintain the arms at a phase difference equivalent to a 50-picometer offset from the dark fringe \cite{Hild:2007DC}. Recently, a new laser amplifier providing 70 Watts of available power, with 50-60 Watts being detected on average at the MPR, was installed in GEO600 \cite{LoughPrivate}.~Thus, we use 50 Watts of laser power for our model of GEO600 as a confident lower bound of the amount of power able to enter the interferometer's dual-recycled cavity through the MPR. The sky-averaged design sensitivity for Advanced LIGO was computed directly using GWINC \cite{2020ascl.soft07020R} rather than Finesse.

An important noise source under active research investigation at high frequencies is the parametric instability of the resonant eigenmodes in the substrate of the test masses in the tens of kHz \cite{parametricInstability, PIobserved}.~The temperature dependence of the test mass eigenmode frequencies renders the parametric instability a difficult noise source to attenuate \cite{PhysRevD.103.022003}.~We did not model this effect in our simulation of LIGO's GW response, using the aLIGO design sensitivity only. In addition, the current configuration of LIGO cannot search for GWs over 10 kHz due to the DC electronic and anti-aliasing (AA) filters suppressing the signal data. Therefore, the sensitivity function we implement for aLIGO represents an optimistic estimate for the future capabilities of this detector design. The resonant eigenmodes of the test masses' substrates have not been observed during observation at GEO600; without high-finesse Fabry-Perot cavities, the circulating laser power in the dual-recycled cavity is not amplified sufficiently to excite these instabilities \cite{GURKOVSKY2007177}, and thus we leave the inclusion of this noise source for future work.~Therefore, we did not simulate parametric instability or include it in our noise budget for the interferometer.

\subsection{\label{subsec:Sensitivities}Sensitivities}

\begin{figure}[t]
    \centering
    \includegraphics[width=\columnwidth]{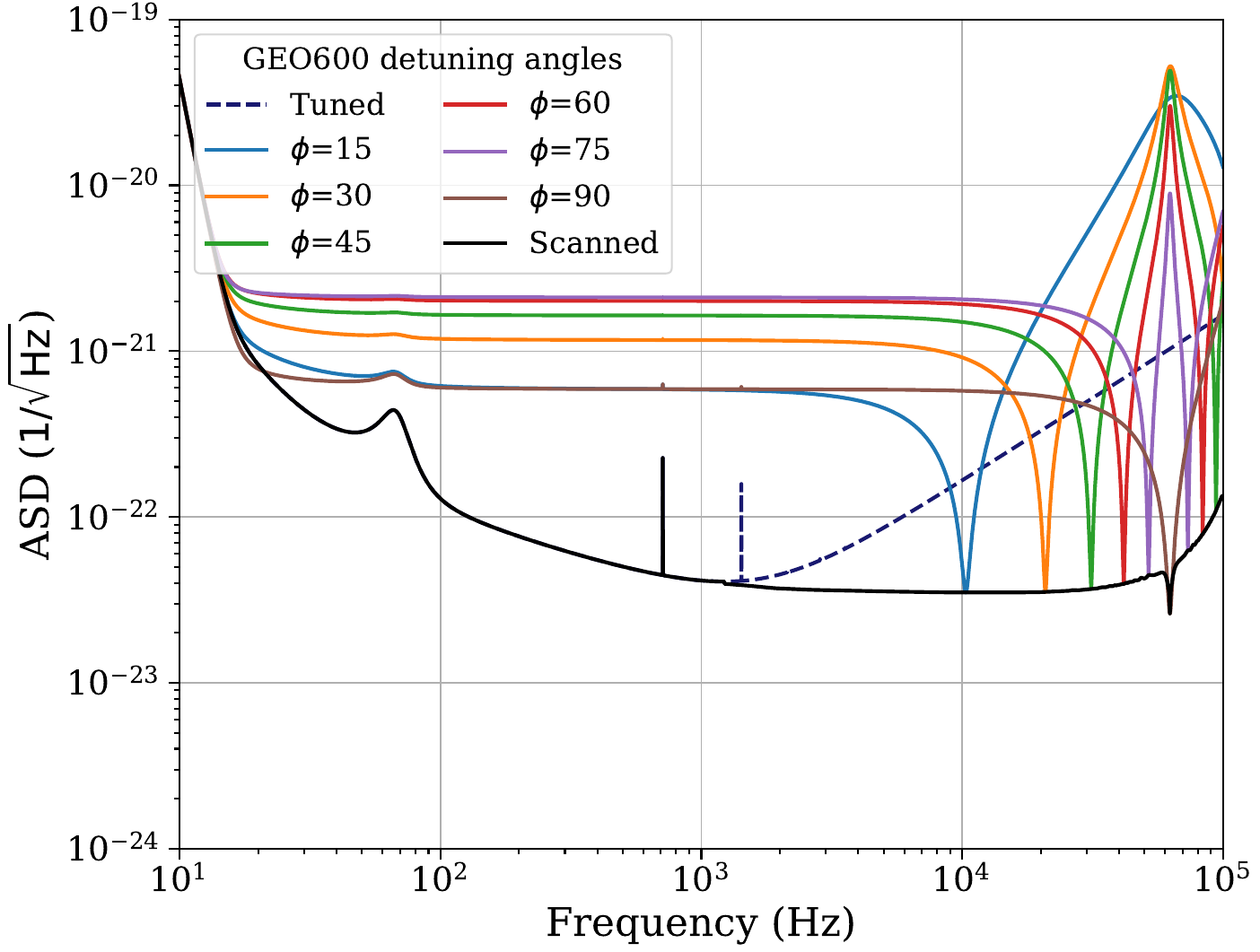}
    \caption{Strain sensitivity $ A_h(f) $ of GEO600 at different detuning angles of the MSR, as well as a scanned sensitivity curve (black), which is the pointwise optimization over MSR detuning angles. As the detuning angle of the MSR is increased from $\phi=0$ (tuned configuration for GEO600), the detection region becomes increasingly narrow-band, in that the frequency bandwidth stays constant while the SRC resonant frequency increases. The scanned sensitivity curve is plotted to show the possible values that can be achieved across the kHz frequency range by shifting the MSR. Note that all strain sensitivities assume a GW sky location of $\theta_{\hat n} =15$ and $\phi_{\hat n}=0$. \href{https://github.com/Chris19j/GEO600-High-Frequency-Modeling/blob/main/FIG2-GEO-detuning.ipynb}{\faFileCodeO}
    }
    \label{GEO-detuning-angles}
\end{figure}
\begin{figure}[h]
    \centering
    \includegraphics[width=\columnwidth]{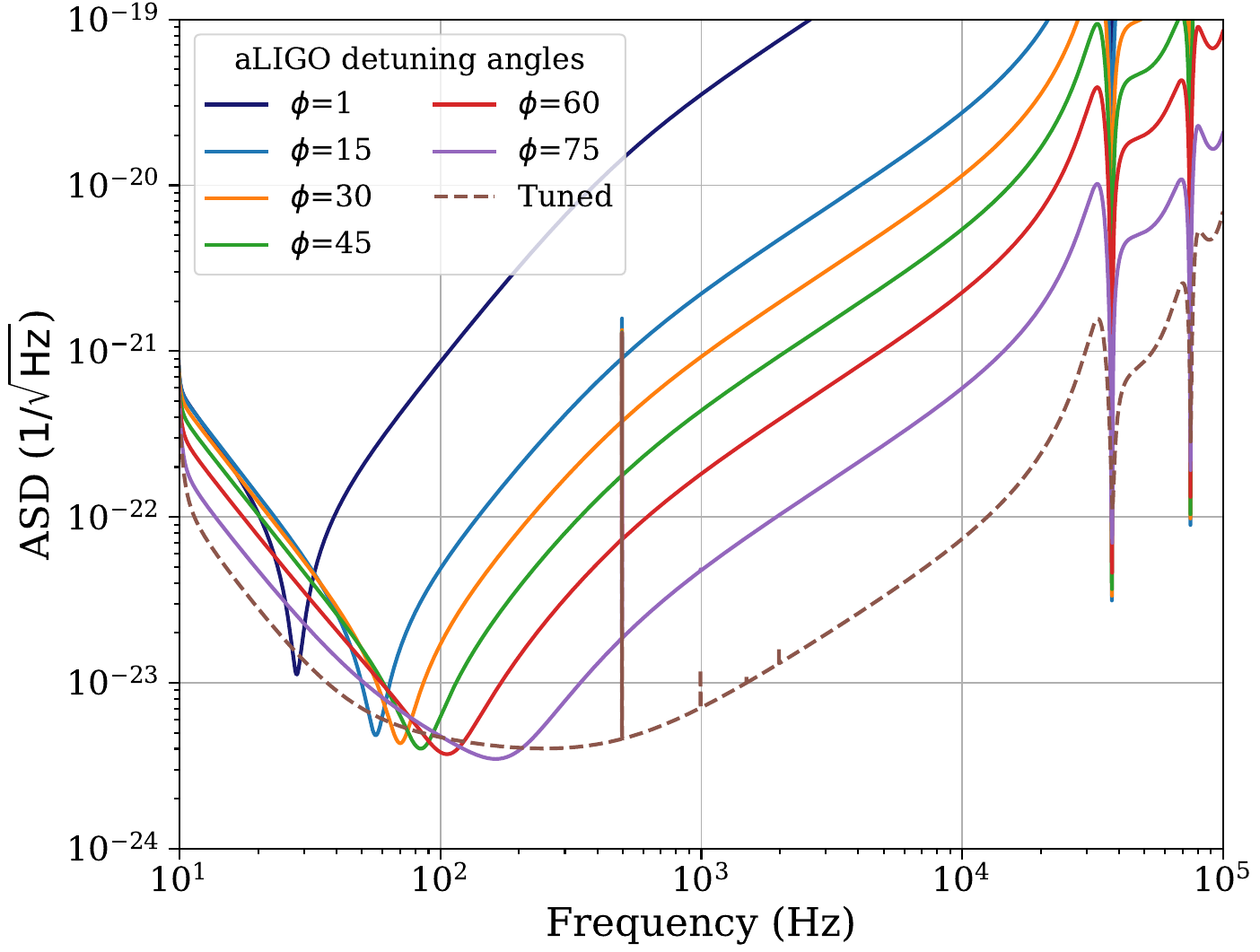}
    \caption{Strain sensitivity  $ A_h(f) $ of the aLIGO design at different detuning angles of the MSR. When the angle of the MSR is decreased from $\phi=90$ (tuned configuration for aLIGO that corresponds to `signal extraction' \cite{Buonanno:2001cj, Heinzel:1999rk}), a region of strong sensitivity gradually shifts to lower frequencies in the neighborhood of $f\sim \mathcal{O}(100~\text{Hz})$, while its bandwidth simultaneously narrows. Note that we use $\phi=1$ to represent the `signal recycling' limit, as setting $\phi=0$ exactly corresponds to a configuration completely dominated by classical noise. There is no enhancement in sensitivity at frequencies in the kHz from aLIGO detuning the MSR. The peaks in sensitivity occurring at integer multiples of $37.5$ kHz (FSR) are created by the Fabry-Perot cavities and are independent of the MSR detuning angle. Note that in this plot we are using current LIGO mirrors --- it is possible that changing the MSR or TM transmissivities then LIGO could have improved high-frequency sensitivity. Note that all strain sensitivities assume a GW sky location of $\theta_{\hat n} =15$ and $\phi_{\hat n}=0$. \href{https://github.com/Chris19j/GEO600-High-Frequency-Modeling/blob/main/FIG3-aLIGO-detuning.ipynb}{\faFileCodeO}
    }
    \label{aLIGO-detuning-angles}
\end{figure}

\begin{figure}[h]
    \centering
    \includegraphics[width=\columnwidth]{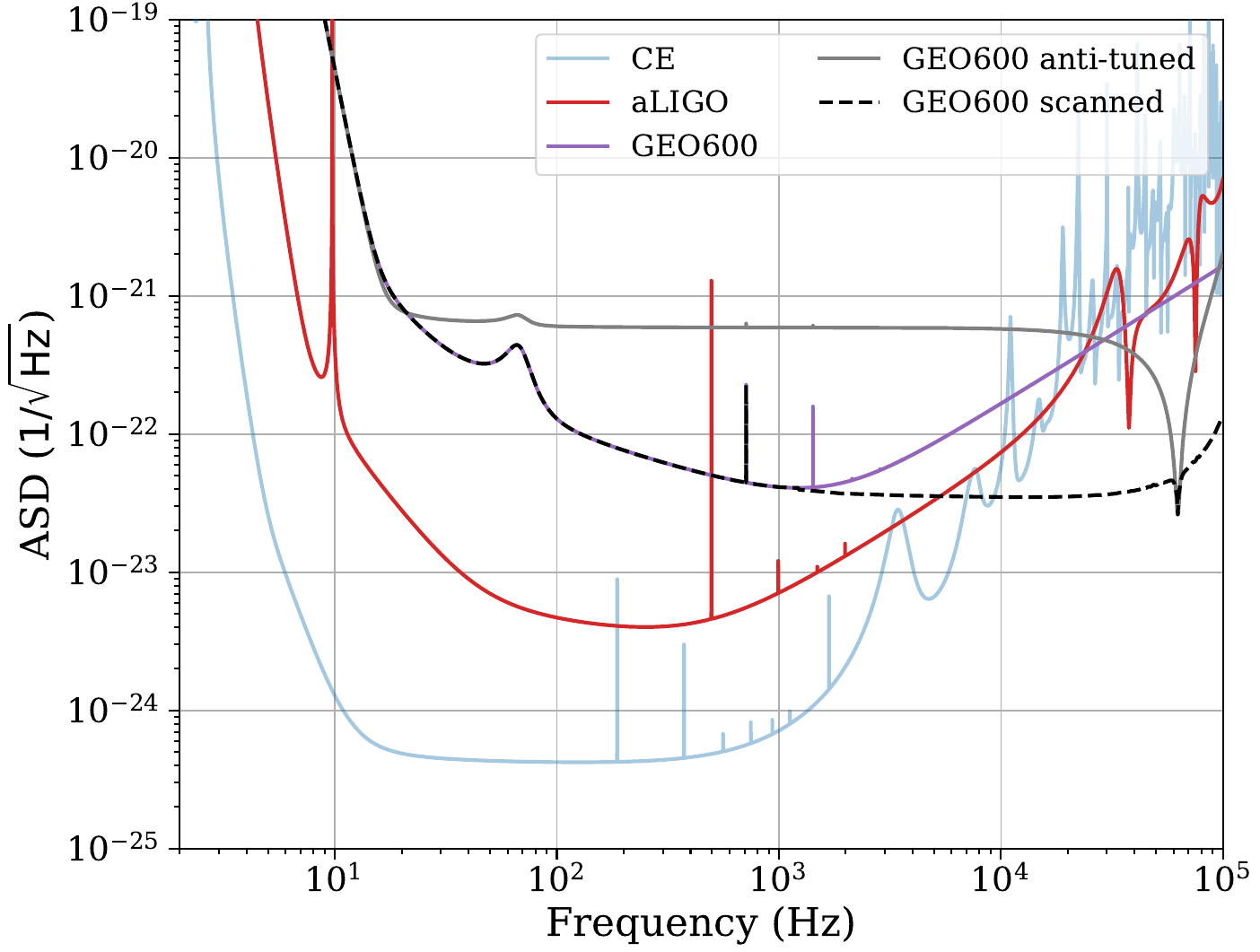}
    \caption{Strain sensitivity $ A_h(f) $ of GEO600 tuned (purple), anti-tuned (gray), and scanned (dashed), which is a summation of all possible MSR detuning angles. The aLIGO design sensitivity (red) \cite{2020ascl.soft07020R}~and Cosmic Explorer's design sensitivity (light blue) \cite{2020ascl.soft07020R} are also shown. GEO600's scanned sensitivity curve shows the pointwise optimization in frequency space over MSR detuning angles. Note that all strain sensitivities assume a GW sky location of $\theta_{\hat n} =15$ and $\phi_{\hat n}=0$. \href{https://github.com/Chris19j/GEO600-High-Frequency-Modeling/blob/main/FIG4-interferometer-sensitivities.ipynb}{\faFileCodeO}
    }
    \label{strain-sensitivity-plot}
\end{figure}

In Fig.~\ref{GEO-detuning-angles}, we show the resulting sensitivity curves for GEO600 with various detuning angles of the MSR, along with a ``scanned" mode where the mirror is swept through the full range of detuning angles and the optimal sensitivity is selected at each value of the detuning. Conversely, the inability of the aLIGO design to scan the interferometer's fundamental resonance peak over a large frequency range is shown in Fig.~\ref{aLIGO-detuning-angles}. The strain sensitivity for the tuned, anti-tuned, and scanned detuning angles of GEO600 is shown in Fig.~\ref{strain-sensitivity-plot}, along with the design sensitivity curves for aLIGO and CE. GEO600 is can achieve higher sensitivity than the aLIGO design starting at $\sim6$ kHz by using various detuning angles to have superior narrow band detection. For our prospective detection calculations, we use individual detuning angles of the MSR with a detection time of one week. The scanned sensitivity curve is only plotted to illustrate the possible values that can be achieved across the kHz frequency range by shifting the MSR.

Fig.~\ref{aLIGO-detuning-angles} portrays how aLIGO experiences resonant amplification in sensitivity at $37.5$ and $75$ kHz. At these specific frequencies, which are integer multiples of the Fabry-Perot cavity FSR, the sideband fields are amplified in power as they resonate in each arm's Fabry-Perot cavity. Detuning the MSR does not affect these resonant frequencies, rendering the sharp features in the aLIGO sensitivity curves at $37.5$ and $75$ kHz independent of the MSR detuning angle.

To assess the potential advantages of using GEO600 as a detection instrument for each source, we compared the SNR of the aLIGO design, GEO600, and GEO600 detuned to various frequencies. We include sky averaging in SNR calculations to account for the reduced sensitivity at certain GW incidence angles where the detector response approaches zero, as shown in FIG.~\ref{Fig.7}. In App.~\ref{Appendix-A}, we discuss the sky-averaging in SNR calculations for the high-frequency antenna response of GEO in more detail (c.f.~Fig.~\ref{Fig.7}). The strain sensitivities of these interferometer configurations, as well as that of Cosmic Explorer's \cite{Reitze:2019iox} for further comparison, are shown in Fig.~\ref{strain-sensitivity-plot}. GEO600's scanned sensitivity curve shows the ASD optimized pointwise over MSR detuning angles at every possible frequency in the kHz range.

\section{\label{sec:high-frequency-sources}High Frequency Sources}
In this section, we investigate two classes of hypothesized high-frequency GW sources and compute benchmarks for how well they could be detected with a detuned GEO600 setup.

\subsection{\label{subsec:ultralight-boson-clouds}Ultralight Boson Clouds}
\subsubsection{GW Strain Model}
A number of proposals to the Standard Model contain ultralight scalar and vector bosons that could have observable astrophysical effects. For example, the scalar axion, derived as a solution to the Strong Charge-Parity problem in QCD \cite{Peccei:1977hh,Peccei:1977ur,Wilczek:1977pj,Weinberg:1977ma}, or the vector dark photon, derived from compactifications of string theory \cite{Goodsell:2009xc, Petraki:2013wwa,Fabbrichesi:2020wbt}.~If the Compton wavelength of the bosons is comparable to the BH mass, the phenomenon of superradiance predicts scenarios with exponential growth of the bosons occurs around spinning BHs, resulting in a macroscopic boson cloud. As the cloud forms, the BH loses angular momentum, causing the emission of continuous GWs at a frequency proportional to the mass of the ultralight bosons \cite{Isi:2018pzk}. 

The gravitational potential of a BH allows for particle bound states that can be approximated by hydrogenic wave functions with radial, orbital, and azimuthal quantum numbers $(n,l, m)$ \cite{PhysRevD.102.063020}.~Thus, the resulting boson cloud formed by superradiance can be described as a hydrogenic wave function, with all of the bosons occupying the same state. For our simulation, we only consider the fundamental radial quantum number $n=0$ and the dominant unstable mode $j=l+s=m=1$—where $l$, $s$, and $m$ are the orbital angular momentum, spin angular momentum, and magnetic quantum numbers, respectively. This unstable mode is chosen because it has the fastest superradiant growth rate in a given system \cite{Isi:2018pzk, PhysRevD.96.024004}, making it the most dominant energy mode. Notably, this also requires that we assume that the BHs which source these scalar axion clouds are \textit{not} primordial in origin, as previous works considering primordial BHs have found that only states with quantum numbers $n\geq5$ can still exist and radiate GWs in the present day \cite{andrewsGeniusBreakoutPaper}. We furthermore constrain our BHs to have initial dimensionless spin parameter of $\chi_\text{i} = 0.7$.

\begin{figure*}[t]
    \centering
    \includegraphics[width=0.495\textwidth]{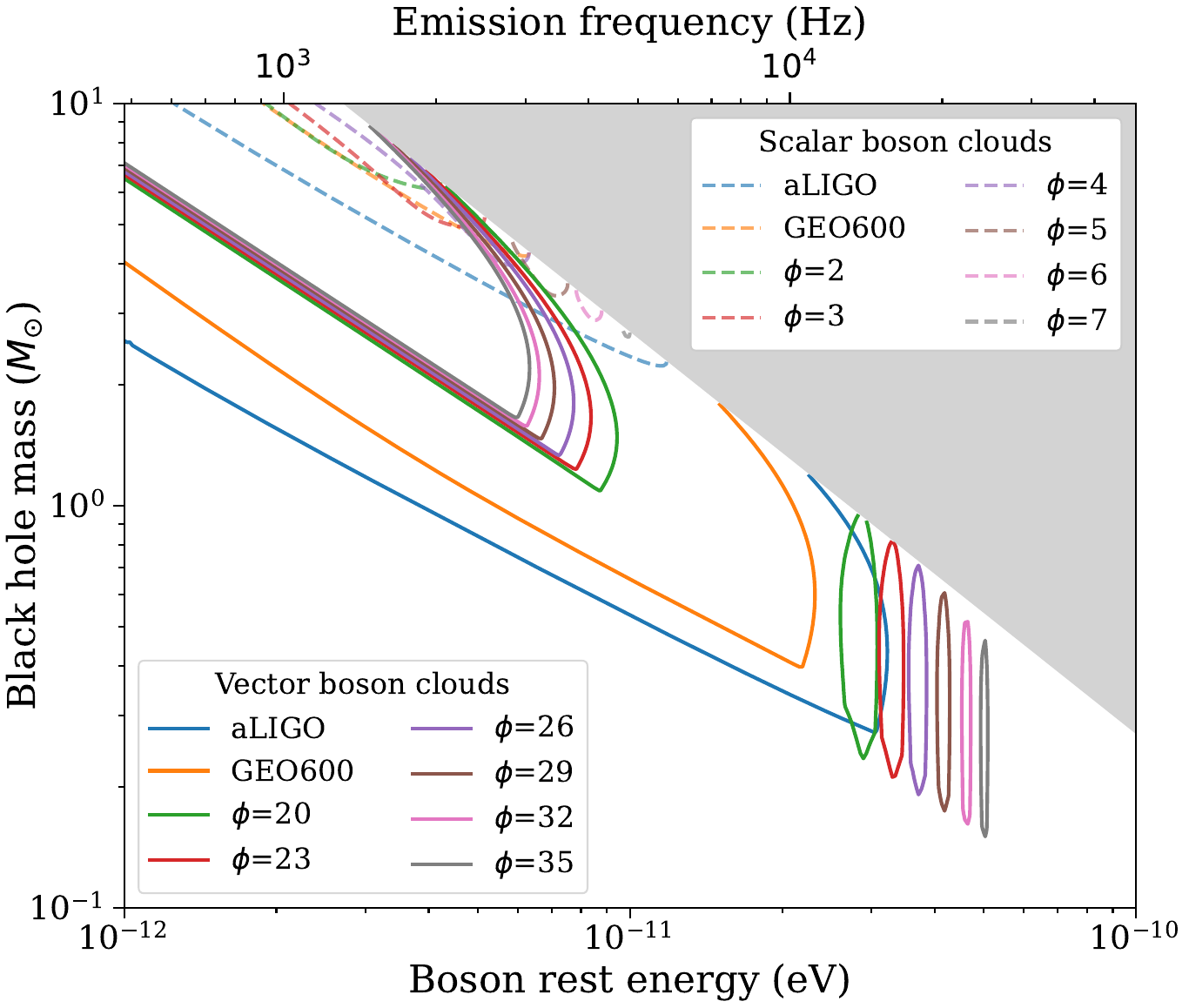}
    \hfill
    \includegraphics[width=0.495\textwidth]{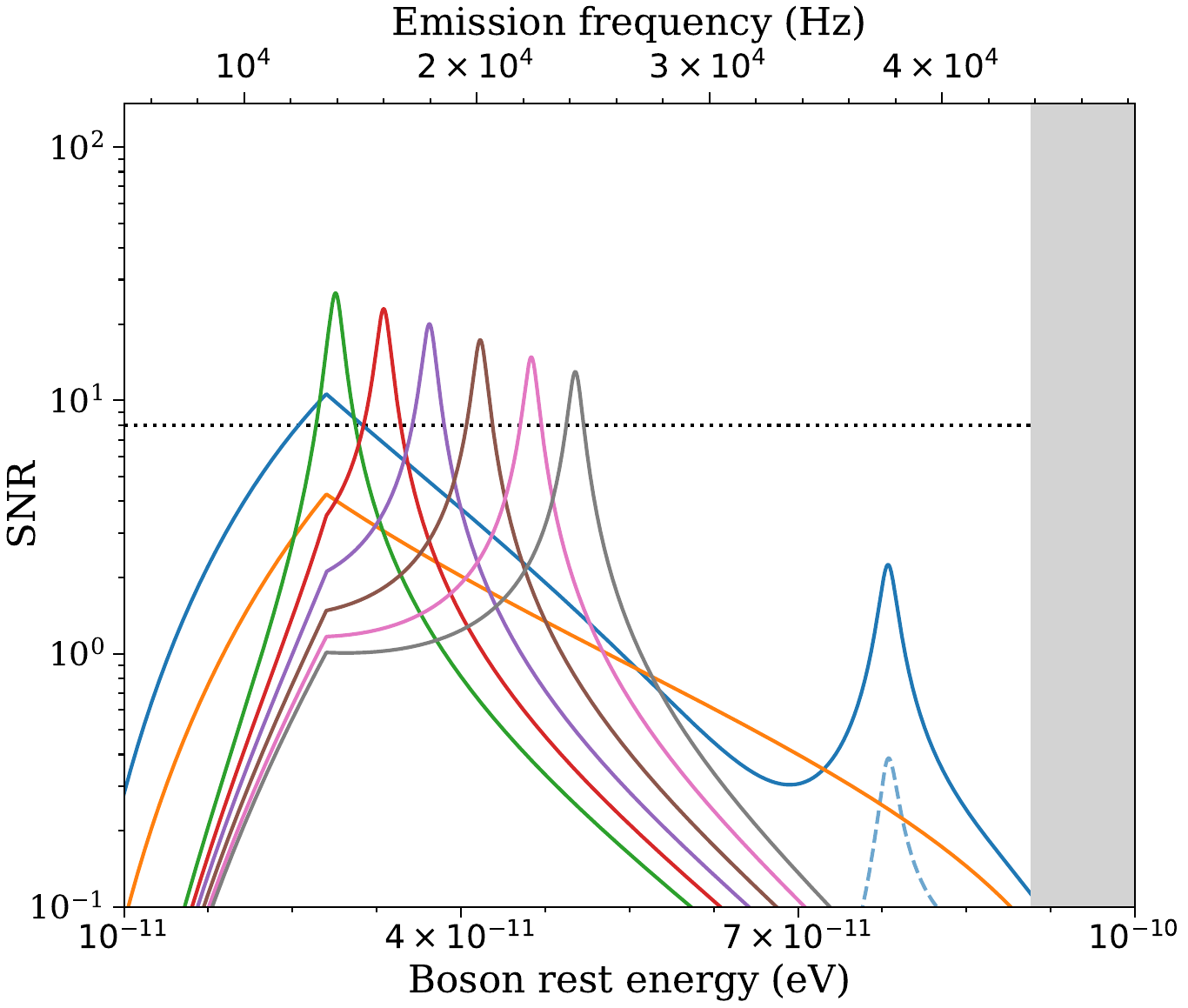}
    \caption{\textbf{Left:} contour lines corresponding to a SNR of eight for GWs sourced from both vector (solid) and scalar (dashed) boson clouds. The contour lines correspond to the strain sensitivities of aLIGO (blue), tuned GEO600 (orange), and GEO600 with different MSR detuning angles. The shaded gray region corresponds to the parameter space where the superradiance condition fails. An SNR of eight can be achieved at frequencies as large as $31$ kHz, with the highest frequency obtained at a detuning angle of $\phi = 45\degree$ (not plotted). \href{https://github.com/Chris19j/GEO600-High-Frequency-Modeling/blob/main/FIG5left-boson-cloud-SNR.ipynb}{\faFileCodeO} \textbf{Right:} SNR of GWs sourced from both vector (solid) and scalar (dashed) boson clouds across a range of boson rest energies for a 0.3 $M_\odot$ BH. The SNR curves are calculated with the same strain sensitivities as the left plot and have the same corresponding colors. The dotted black line represents where the SNR is eight. The sharp increase in SNR in aLIGO at 37.5 kHz (FSR) arises due to sideband resonance in its Fabry-Perot cavities, while the kink in the SNR function on the right plot around 14 kHz corresponds to the sudden drop-off in integration time when $\tau_\text{GW}$ falls below one week. All GWs are sourced from within the galaxy, being simulated from a distance of $30$ kPc, and assume a sky-averaged interferometer response. \href{https://github.com/Chris19j/GEO600-High-Frequency-Modeling/blob/main/FIG5right-boson-cloud-SNR.ipynb}{\faFileCodeO}}
    \label{Fig.5}
\end{figure*}

The amplification process can only occur when the angular velocity of the black hole at the event horizon is higher than the angular frequency of the bosons divided by the magnetic quantum number \cite{Isi:2018pzk}:
\begin{equation}
    \omega_\mu/m < \Omega_\mathrm{BH}(\chi_\text{i}),
\end{equation}
where $\omega_\mu =\mu/\hbar$, $m=1$, and $\Omega_\mathrm{BH}(0.7)= 1.38 \times 10^5\! \left(\!\frac{0.3M_\odot}{M}\!\right)\text{s}^{-1}$. Rearranging this superradiance condition constrains the rest energy of bosons surrounding a BH of mass $M$ to be  $\mu < 9.10 \times  10^{-11} \!\left(\!\frac{0.3 M_\odot}{M}\!\right) \text{eV}$. For boson clouds with boson rest energies in the target range of $10^{-12}$ to $10^{-10}$ eV, the GW emission frequency is given by \cite{Isi:2018pzk}~
\begin{equation}
    f_\text{GW} \approx 50 \, \text{kHz} \times \frac{\mu}{10^{-10}\text{ eV}}.
\end{equation}

Since there is an upper bound on $\mu$ based on the corresponding mass of the BH it is orbiting, there is also an upper bound on the GW emission frequencies that are possible for every BH mass. The GW strain for vector bosons is given as
\begin{equation}
    \hspace{-0.76em}
    h_0^v \approx 10^{-22} \!\left(\!\frac{M}{0.3 M_\odot}\!\right)\! \!\left(\frac{\alpha}{0.2}\right)^{\!5} \!\!\left(\!\frac{0.03 \, \text{Mpc}}{r}\!\right)\! \!\left(\!\frac{0.7 - \chi_f}{0.1}\!\right)\!,
\end{equation}
where $\alpha$ is the ratio of the characteristic lengths of the boson cloud and the BH, and $\chi_\text{f}$ is the BH spin after superradiance occurs. $\alpha$ and $\chi_\text{f}$ are given as
\begin{equation}
    \alpha = \frac{\mu GM}{\hbar c^3}, \quad \chi_\text{f} = \frac{4\alpha_f}{4\alpha_f^2 + 1}.
\end{equation}
Here, $\alpha_f = \frac{\mu GM_f}{\hbar c^3}$, and $M_f=0.9M$ is the mass of the BH after the boson cloud has extracted its energy during the superradiant growth \cite{Isi:2018pzk}.~The GW signal duration for vector bosons is given as
\begin{equation}
    \tau_{\text{GW}}^v \approx 1.81 \, \text{s} \left(\frac{M}{0.3 M_\odot}\right) \left(\frac{0.2}{\alpha}\right)^{11}.
\end{equation}
For scalar bosons, the GW strain and signal duration are both given as
\begin{equation}
    \hspace{-0.9em}
    h_0^s \approx 10^{-25} \!\left(\!\frac{M}{0.3 M_\odot}\!\right) \!\!\left(\frac{\alpha}{0.2}\right)^{\!7} \!\!\left(\!\frac{0.03 \, \text{Mpc}}{r}\!\right) \!\!\left(\!\frac{0.7 - \chi_f}{0.1}\!\right)\!,
\end{equation}
\begin{equation}
    \tau_{\text{GW}}^s \approx 31 \, \text{days}\left(\frac{M}{0.3 M_\odot}\right) \left(\frac{0.2}{\alpha}\right)^{15}.
\end{equation}

\subsubsection{Detection Prospects}

Our figure of merit to compare the detection capabilities of our detuned GEO600 configurations against LIGO is the sky-averaged SNR produced by a characteristic high-frequency source in each detector. Since the boson clouds produce a monochromatic signal, the sky-averaged SNR is given by \cite{Maggiore:2008ulw}
\begin{equation}
    \langle \rho^2\rangle = \frac{|h_0^b|^2 T}{ \langle S_h(f_0) \rangle },
\end{equation}
where $b = s,v$ for either scalar or vector boson strain and $h_0$ is averaged over source inclination angles. Here, $T\equiv\text{min}\{1\text{ week},\tau_{GW}\}$ is the observation time. Note that $\langle S_h(f_0) \rangle = \langle A_h(f_0)^2 \rangle =  \left(\!\frac{A_h(f_0)}{\mathcal{R}(f_0)}\!\right)^2$.

In Fig. \ref{Fig.5}, we show contours of SNR equal to 8 for vector and scalar boson clouds for different interferometer configurations, representing regions of this parameter space which are detectable by each system. The plot shows the feasibility of certain configurations of GEO600 detecting GWs for different combinations of boson rest energies and BH mass. We also present a horizontal cross-section of the contour plot, which shows the SNR of the interferometer at different boson rest energies for a given BH mass $m_{\text{BH}}=0.3M_\odot$.

Both the tuned GEO600 and aLIGO would be able to detect vector boson clouds over a large parameter space, though aLIGO generally tends to outperform GEO600 due to its superior broadband sensitivity. However, over a range of boson rest masses (and therefore characteristic GW frequencies), GEO600 can gain a competitive advantage over aLIGO in a narrow band around that characteristic frequency with a careful choice of the MSR detuning angle. Indeed, at frequencies above $\sim6.5$ kHz, there exists some detuned configuration of GEO600 which can attain greater SNR for boson cloud GWs than aLIGO. 

Furthermore, for frequencies $\gtrsim15$ kHz, GEO600 obtains an SNR of eight with a week of integration time while aLIGO is not able to reach this threshold. By detuning the MSR and observing at that frequency range for a week, GEO600 can search for GWs sourced from boson clouds from $500$ Hz to $31.3$ kHz, with the latter frequency being obtained with a detuning angle of $\phi = 45\degree$. Note that the pattern of high-frequency narrow bands in Fig.~\ref{Fig.5} is only shown up to $\phi=35\degree$, which results in a detection bandwidth centered at $24.5$ kHz. 

Thus, GEO600 would be able to detect GWs from a larger frequency range, increasing the likelihood the interferometer observes the phenomena. It would be advantageous to use GEO600 as a high-frequency GW interferometer targeting GWs sourced from boson clouds within the galaxy in comparison to using aLIGO. Expanding the region of the BH-boson cloud system parameter space which sources detectable GWs increases the likelihood of detecting the superradiance phenomenon. Note that the current operational LIGO design would not be able to detect GWs over $10$ kHz due to the DC electronic and anti-aliasing (AA) filters suppressing the signal data past this cutoff frequency.

Conversely, Fig.~\ref{Fig.5} shows that both interferometers have a very small area of parameters that result in the possible detection of \textit{scalar boson clouds}, and the area does not extend into the high-frequency emission range like the vector boson parameter area does. The primary cause for the inability to detect high-frequency GWs from scalar boson clouds is that the strain is a factor of $\sim10^3 $ weaker that of vector boson clouds. 

\subsection{\label{subsec:ssm-pbh}Sub-Solar Mass Compact Object Mergers}

\subsubsection{GW Strain Model}

Sub-solar mass (SSM) compact objects are a compelling category of potential GW sources in the high-frequency regime accessible to current and future detectors. While conventional astrophysical compact objects such as neutron stars and black holes are typically formed with masses above $M_\odot$, various theoretical models predict the existence of SSM compact objects through unconventional formation channels or as entirely new types of exotic matter configurations \cite{Aggarwal:2020olq}.~Outside the conventional BH-formation theory of stellar evolution, PBHs may have formed from the gravitational collapse of density perturbations in the early Universe \cite{1974MNRAS.168..399C, Zeldovich:1967lct, Hawking:1971ei},~and BHs formed from the gravitational collapse of dark matter halos \cite{darkhaloBlackHoles}~have been proposed as SSM compact objects.~Additionally, there exist theories of SSM exotic compact objects, such as gravitino stars \cite{gravitinoStars}, boson stars \cite{bosonStar}, moduli stars \cite{moduliStars}, and gravastars \cite{gravastars}. If a SSM binary were found, classification between BBH and BNS is possible since tidal deformabilities strongly affect the waveform for light NS \cite{Golomb:2024mmt}.

We simulate GWs from SSM compact binary mergers within the mass range of $10^{-6}~M_{\odot}$ to $1~M_{\odot}$.~Note that the detection of GWs sourced from compact objects within this mass range would not be provide the sufficient parameters to distinguish that the object is a PBH specifically. In addition, if the source were a PBH, a detection within this mass range could only provide evidence that PBHs constitute some fraction of dark matter \cite{Franciolini:2022htd}. We compute the characteristic strain from the binary merger of SSM compact objects using the inspiral, merger, and ringdown waveforms  available in PyCBC's \texttt{IMRPhenomD} package \cite{Biwer_2019}.~For simplicity, we assume an inclination angle of $0\degree$, making $\Tilde{h}_+(f)\!=\!\Tilde{h}_\times(f)$ and $\Tilde{h}(f)=\mathcal{R}(f)\Tilde{h}_0(f)$ (see Appendix~\ref{subsec:Sky-Averaging}).~We calculated the sky-averaged SNR using the chirp formula given by \cite{Maggiore:2008ulw}
\begin{equation}
    \langle \rho^2 \rangle = 4\int_{f_\text{min}}^\infty \frac{|\Tilde{h}(f)|^2}{\langle S_h(f) \rangle}df,
\end{equation}
where $f_\text{min}$ is the GW frequency emitted exactly one week before the compact objects merge \cite{Maggiore:2008ulw}: 
\begin{equation}
    f_{\text{min}}(\tau) \simeq 4.72 \, \text{Hz} \left( \frac{0.087 M_\odot}{M_c} \right)^{5/8} \left( \frac{1 \, \text{wk}}{\tau} \right)^{3/8}, 
\end{equation}
where the chirp mass $M_c$ for two $0.1M_\odot$ BHs is equal to 0.087$M_\odot$, and in general is given by \cite{Maggiore:2008ulw}
\begin{equation}
    M_c = \frac{(m_1 m_2)^{3/5}}{(m_1 + m_2)^{1/5}}\,.
\end{equation} 
In our calculations, we set the time to coalescence $\tau$ equal to one week.

\begin{figure}
    \centering
    \includegraphics[width=\columnwidth]{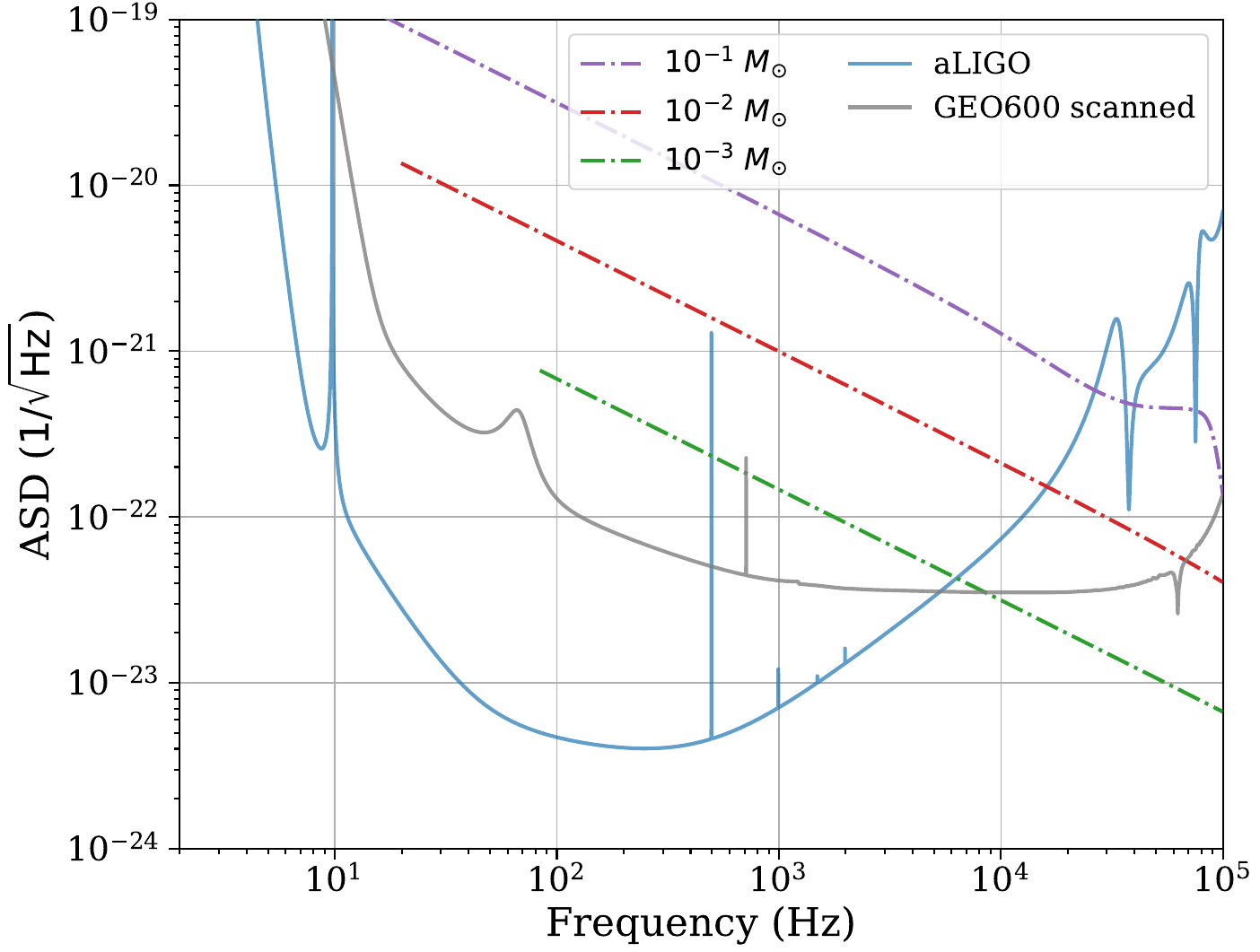}
    \caption{Strain sensitivity $ A_h(f) $ of the aLIGO design (light blue) and scanned GEO600 (light gray). Three different GW waveforms are plotted as $2|\Tilde{h}_+(f)|\sqrt{f}$. The GW amplitudes are from compact binary mergers of three different sub-solar masses, observed from a distance of $30$ kPc. Note that all strain sensitivities assume a GW sky location of $\theta_{\hat n} =15$ and $\phi_{\hat n}=0$. \href{https://github.com/Chris19j/GEO600-High-Frequency-Modeling/blob/main/FIG6-compact-object-waveforms.ipynb}{\faFileCodeO}
    }
    \label{Fig.6}
\end{figure}

\subsubsection{Detection Prospects}
Our figure of merit to compare the detection capabilities of our detuned GEO600 configurations against LIGO is the sky-averaged SNR produced by a characteristic high-frequency SSM compact binary merger in each detector. In Fig.~\ref{Fig.6}, we show the GW strain of compact mergers with individual BH masses of $10^{-1}$, $10^{-2}$, and $10^{-3}~M_{\odot}$, located a distance of 30 kPc from the detector. As the chirp mass decreases, $f_{\text{min}}$ increases while the GW strain is reduced. The frequency-domain signal is calculated with an observation time of one week, and the SNR is related to the area in the plot between the GW waveform $\Tilde{h}(f)$ and the strain sensitivity corresponding to a single interferometer configuration $A_h(f)$. Note that FIG.~\ref{Fig.6} only shows the scanned sensitivity of GEO600, which is the ASD optimized pointwise in frequency space over MSR detuning angles at all frequencies and not a realizable configuration itself.

We find that there are not SSM binaries which can be detected easier with a detuned GEO600 detector. Over the frequency range where the detuning-scanned GEO600 has a better sensitivity than aLIGO, the strain from the merger is weak, and thus the accumulated SNR from any single detuned configuration is so little that it does not sway the total SNR in favor of GEO600. While the time domain signal is stronger at high frequencies, the early inspiral dominates the SNR because it has more cycles in band. If there were a SSM binary that were very slowly evolving in frequency at a timescale of $\sim$ months, then detuning GEO to the exact frequency could improve the sensitivity compared with LIGO. For nearly monochromatic SSM signals, one would see a similar improvement as to what we saw in Sec.~\ref{subsec:ultralight-boson-clouds}. Thus, GEO600 is not an optimal ground-based interferometer for the detection of most SSM compact binary mergers.

\section{\label{sec:conclusion}Conclusion}
In this work, we investigated how the high-frequency sensitivity of GEO600 can be scanned by detuning the location of the MSR. The MSR creates an optical cavity that enables narrow-band resonant amplification of GW-driven sideband fields. Detuning the MSR changes which frequencies satisfy the resonant condition, allowing GEO600 to scan a resonant feature with a bandwidth of $\sim 2$ kHz over a range of high frequencies. We estimated the SNR in the experiment for characteristic high-frequency GW sources located within the Milky Way. This setup of detuning GEO600 does not result in an improvement in searches for SSM binaries due to their long inspiral in low frequencies, and the strain from the merger being weak at high frequencies. However, GEO600 could outperform the aLIGO design sensitivity for the detection of GWs sourced from nearly monochromatic sources like boson clouds within the galaxy. By specifying a MSR detuning angle between $0\degree$ and $45\degree$, GEO600 would be able to detect GWs sourced from vector boson clouds at any $\sim 2$ kHz band of frequencies between $500$ Hz to $31.3$ kHz, gaining the advantage over the aLIGO design sensitivity at $15.1$ kHz and over O4 LIGO at $10$ kHz.

In the future, it would be interesting to extend our analysis on GEO600 to include time domain simulations as there are lots of complications associated with maintaining the resonance condition of the MSR while it is detuned for high-frequency GW detection. Neural-network–based sensing and auxiliary sub-carrier field injections are innovative control system methods to maintain the detuned MSR in the resonant state during high-frequency detection \cite{neuralNetworkSensing, MSRcontrol}. In addition, a new FPGA-based fast data acquisition system has been installed that samples data up to $100$ MHz, extending data collection availability into the hundreds of kHz \cite{LoughHFpaper}. Throughout this work, we used frequency-independent squeezing where we optimized the squeezing angle to maximize the sensitivity at the resonant frequencies. Normally for a Michelson interferometer, a filter cavity is designed to rotate the light in a frequency-dependent way which results in higher bandwidth for the detuned detector. This is a very interesting followup work to investigate possible ways to achieve frequency-dependent squeezing that is also tunable \cite{atomicSqueezing, atomicSqueezing2, quantTeleSqueez, EPRsqueezing}. We leave these studies to future work.

\begin{acknowledgments}
C.J.~acknowledges the support of the National Science Foundation Research Experience for Undergraduates program (NSF REU), the LIGO Laboratory Summer Undergraduate Research Fellowship program (LIGO SURF), and the California Institute of Technology Student-Faculty Programs (Caltech SFP). A.L.~is supported by the John and Fannie Hertz Foundation. B.S.~acknowledges support by the National Science Foundation Graduate Research Fellowship under Grant No.~DGE-1745301. Y.C.’s research is supported by the Simons Foundation (Award No.~568762), the Brinson Foundation, and the National Science Foundation (via Grants No.~PHY-2309211 and PHY-2309231). We wish to acknowledge Aaron Goodwin-Jones, Su Direkci, and James Gardner for their useful discussion and comments on the manuscript, and James Lough for his useful information about GEO600's parameters.
\end{acknowledgments}

\appendix

\section{Antenna Response Patterns} \label{Appendix-A}
In this appendix, we provide a derivation for the sky-averaged antenna response pattern at high frequencies for two interferometer geometries.

\subsection{Unfolded Michelson Interferometer}
When studying high-frequency $(f\sim c/L_\mathrm{arm})$ interferometric detection, an important piece of physics which typically can be neglected in low-frequency calculations is the evolution of the GW strain as a photon travels out and back along an arm of the interferometer. For certain frequencies and angles of incidence, the phase accumulated by a photon during a full round-trip in the interferometer can deviate significantly from $2hk_\mathrm{photon}L$ --- at some frequencies even being identically zero. Thus, to properly model the response of GEO600 and LIGO to high-frequency GWs, we must compute the generalized response function for a folded and unfolded Michelson interferometer which includes the GW and photon propagation effects. This is done originally in Refs.~\cite{PhysRevD.96.084004,Schilling:1997id,Rakhmanov_2008}

We will now replicate the derivation of the high-frequency response of a Michelson interferometer as outlined in \cite{Rakhmanov_2008}. The time-domain form of a plane wave GW is given by 
\begin{equation}
    h_{i j}(t, \vec{x})=h_{+}(t, \vec{x}) e_{i j}^{+}(\hat{n})+h_{\times}(t, \vec{x}) e_{i j}^{\times}(\hat{n})\,,
\end{equation}
where $\hat n$ points toward the GW source. The polarization tensors are defined via 
\begin{subequations}
    \begin{equation}\label{C6a}
        \hat e^+_{ij}=\ell_i \ell_j- m_i m_j,
    \end{equation}
    \begin{equation}\label{C6b}
        e^\times_{ij}=\ell_i m_j+\ell_i m_j,
    \end{equation}
\end{subequations}
where the unit vectors $\hat \ell,\hat m,\hat n$ form a right-handed orthonormal basis. In this discussion, we suppress the relative rotation that is due to the polarization angle. In the long-wavelength approximation, the Michelson interferometer response is given by
\begin{equation}
    h(t)=\frac{1}{2}\left(a^i a^j-b^i b^j\right) h_{i j}(t, \vec{0})\,.
\end{equation}
One can then write this as 
\begin{equation}
    h(t)= F_+(\hat n) h_+(t) + F_\times(\hat n) h_\times(t) \,.
\end{equation}
If we take unit vectors $a^i$ and $b^i$ to represent the two arms of the detector, the antenna patterns are given by 
\begin{equation}\label{eq:DCpatterns}
    F_A(\hat{n})=\frac{1}{2}\left(a_i a_j-b_i b_j\right) e_A^{i j}(\hat{n}) \, ,
\end{equation}
where we use $A=+,\times$ to denote polarization. In this long-wavelength regime, the antenna patterns do not depend on the frequency.

Let us now generalize the results of Eq.~\eqref{eq:DCpatterns} to high-frequency regime. As discussed in Eqs.(15-17) of Ref.~\cite{Rakhmanov_2008}, the response of the GW detector is related to the time delay accumulated in the round trip
\begin{equation}
    h(t) = \frac{1}{2 T} \left[\delta T_{\mathrm{r.t.},\hat a}^\mathrm{mich}(t)-\delta T_{\mathrm{r.t.},\hat b}^\mathrm{mich}\right]
\end{equation}
where $\delta T_{\mathrm{r.t.},\hat a}^\mathrm{mich}$ is the round trip time delay incurred from the presence of the GW along the $\hat a$ direction and $T=L/c$. The round trip time delay along direction $\hat a$ is the sum of the delay from the out and back journey\footnote{Note that the normalization $1/2T$ comes from enforcing that $\max_{\hat n} F_A(\hat n,f) = 1$ when $f=0$.}
\begin{equation}
    \delta T_{\mathrm{r.t.},\hat a}^\mathrm{mich}(t)=\delta T_{\hat a}(t-T)+\delta T_{\hat a}^{\prime}(t)\,,
\end{equation}
where 
\begin{subequations}
    \begin{equation}
        \delta T_{\hat a}(t)=\frac{1}{2 c} a^i a^j \int_0^L h_{i j}\left(t-T+\frac{\xi}{c}+\frac{\hat{n} \cdot \hat{a}}{c} \xi\right) \mathrm{d} \xi\,,
    \end{equation}
    \begin{equation}
        \delta T_{\hat a}^{\prime}(t)=\frac{1}{2 c} a^i a^j \int_0^L h_{i j}\left(t-\frac{\xi}{c}+\frac{\hat{n} \cdot \hat{a}}{c} \xi\right) \mathrm{d} \xi\,.
    \end{equation}
\end{subequations}
We note that the time translation property of the GW is used $h_{ij}(t, \vec{x})=h_{ij}(t+\vec{x} \cdot \hat{n} / c)$, which is why these expressions only are a function of one number.

Using the expressions for the round trip time delay, we can write the frequency-dependent antenna patterns in the frequency domain. In the Fourier domain, the time delay is 
\begin{equation}
    \frac{\delta \tilde T_{\mathrm{r.t.},\hat a}^\mathrm{mich}(f)}{T}=a_i a_j D(\hat{a}, f) e_A^{i j}(\hat{n}) \tilde{h}_A(f)\, ,
\end{equation}
where we define the transfer function $D(\hat a,f)$ as 
\begin{equation}
    \hspace{-1em}
    D(\hat{a}, f)=\frac{\mathrm{e}^{-\mathrm{i} 2 \pi f T}}{2}\!\left[\mathrm{e}^{\mathrm{i} \pi f T_{+}} \operatorname{sinc}\!\left(\pi f T_{-}\right)+\mathrm{e}^{-\mathrm{i} \pi f T_{-}} \operatorname{sinc}\!\left(\pi f T_{+}\right)\right],
\end{equation}
and define $T_\pm = T(1\pm \hat a \cdot \hat n)$. In the end, the frequency-domain GW signal is 
\begin{equation}
    \tilde h(f) = F_+(\hat n,f)\tilde h_+(f) + F_\times(\hat n,f) \tilde h_\times(f)\,,
\end{equation}
where the frequency-dependent antenna patterns are defined via \cite{Rakhmanov_2008} as
\begin{equation}
    F_A(\hat n,f) = \frac{1}{2}\left[a_i a_j D(\hat{a}, f)-b_i b_j D(\hat{b}, f)\right] e_A^{i j}(\hat{n}) \,.
\end{equation}

\subsection{Folded Michelson Interferometer}

GEO600 has folded interferometer arms, making the antenna patterns deviate at high frequencies from that of a MI with unfolded arms \cite{LoughHFpaper,Schilling:1997id}. To ensure that the SNR of GEO600 is accurate in the high-frequency range of a few kHz to tens of kHz, we will now replicate the derivation of Ref.~\cite{LoughHFpaper} and derive the folded-arm MI antenna response pattern. In this derivation, we will ignore the effects caused by the folding angle of GEO600's interferometer arms. Since $\alpha_\mathrm{fold} =0.42$ mrad \cite{Heinert:2014pba}, the effects of the triangular shape of the fold occur when the GW wavelength approaches $\lambda_\mathrm{gw}\sim\alpha_\mathrm{fold}L\approx 0.25 $ m, which corresponds to a GW frequency of $f\gtrsim 1$ GHz.

One can now do the analysis of the time delay that the light accumulates along the trip of a folded detector in arm $\hat a$. This can be written as
\begin{equation}
    \delta T_{\mathrm{r.t.},\hat a}^\mathrm{fold}(t)=\delta T_{\hat a}(t-3T)+\delta T_{\hat a}^{\prime}(t-2T) +\delta T_{\hat a}(t-T)+\delta T_{\hat a}^{\prime}(t)\,,
\end{equation}
where we use the same definitions as the previous section. It is useful to relate the round trip of a folded Michelson to a unfolded Michelson trip delay:
\begin{equation}
    \delta T_{\mathrm{r.t.},\hat a}^\mathrm{fold}(t) = \delta T_{\mathrm{r.t.},\hat a}^\mathrm{mich}(t-2T)+\delta T_{\mathrm{r.t.},\hat a}^\mathrm{mich}(t) \, .
\end{equation}
If we take the Fourier transform of this, we find that the folded Michelson time delay is equal to 
\begin{equation}
    \delta \tilde T_{\mathrm{r.t.},\hat a}^\mathrm{fold}(f) = \left( 1 + e^{4 \pi i f T} \right)\delta \tilde T_{\mathrm{r.t.},\hat a}^\mathrm{mich}(f) \, .
\end{equation}
Since the round trip time delay is changed by a non-directionally-dependent factor, the antenna pattern of a folded Michelson interferometer is directly proportional to an unfolded Michelson interferometer via
\begin{equation}\label{eq:foldedantenna}
    F_A^\mathrm{fold}(\hat n,f)=\frac{1 + e^{4 \pi i f T}}{2} F_A^\mathrm{mich}(\hat n,f) \, ,
\end{equation}
where the extra factor of 2 comes from the DC antenna pattern normalization. In this equation, we stress that $F_A^\mathrm{mich}(\hat n,f)$ is a Michelson interferometer of length $600$m while $F_A^\mathrm{fold}(\hat n,f)$ is a folded Michelson interferometer of length $600$m $+$ $600$m $=1200$m. So while it is understood that GEO600 has an effective arm length of 1200m because of the folding, the high-frequency antenna patterns of GEO600 behave more like a 600m detector times a factor. One notable feature of the folded detector is that there is zero response for every angle at frequency values of 
\begin{equation}\label{eq:nullfoldedfreq}
    f = \frac{c}{4 L}\left( 1+2n \right)\, , \quad  n\in \mathbb{N}_0\,.
\end{equation}
In the next section, we will see that there are complete dips in the angle averaged sensitivity for a folded detector.

\begin{figure*}[t]
    \centering
    \includegraphics[width=\columnwidth]{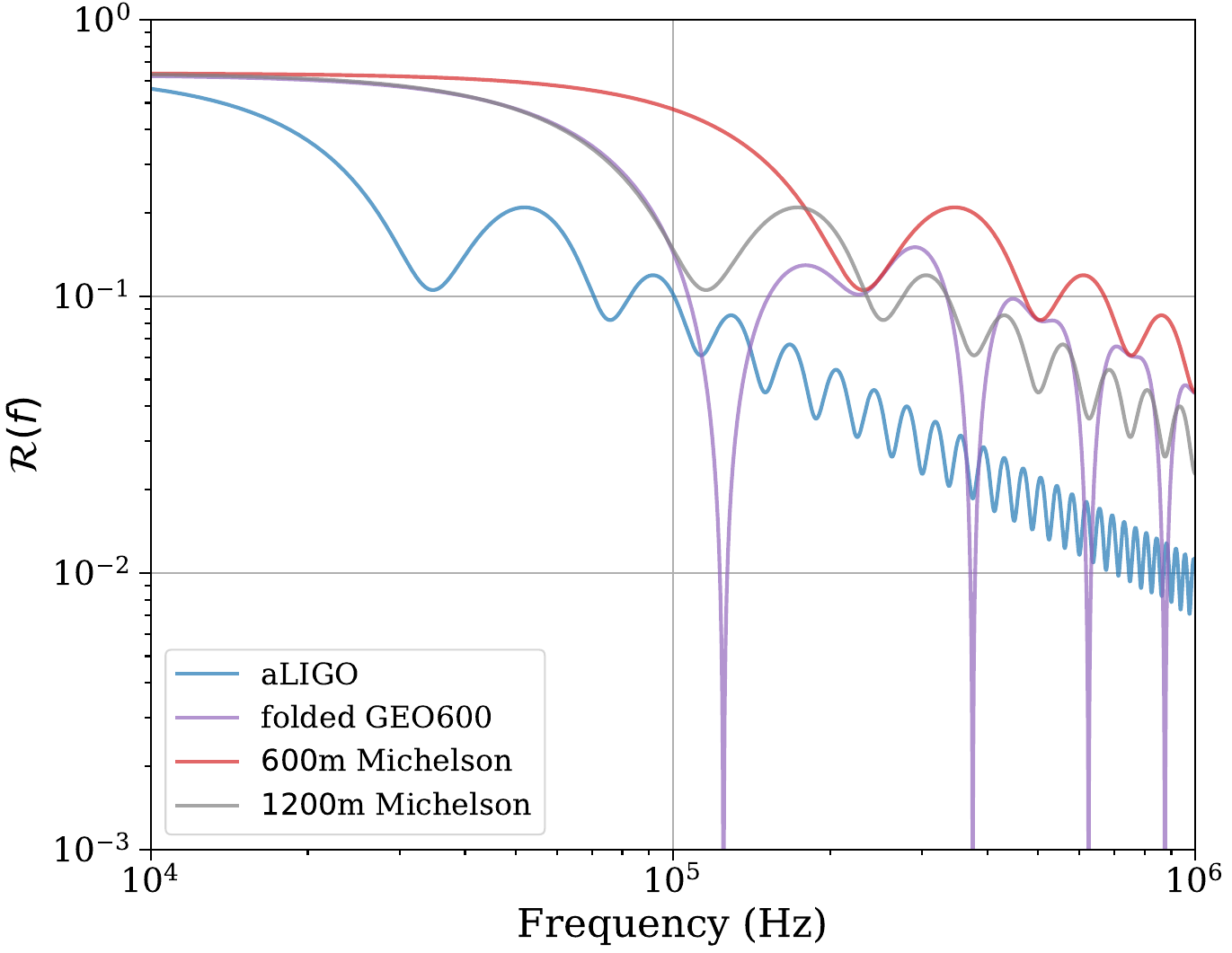}
    \includegraphics[width=\columnwidth]{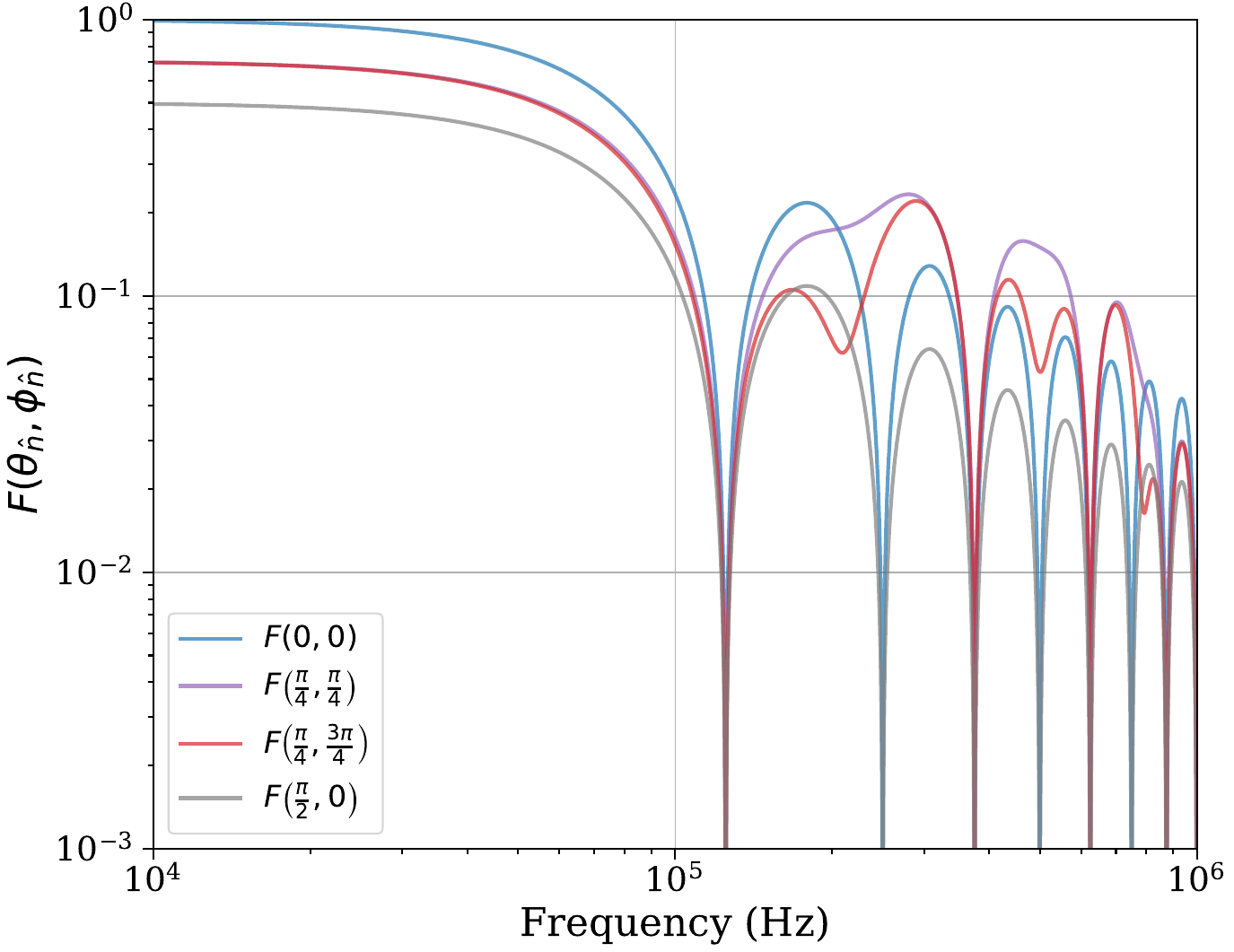}
    \caption{\textbf{Left:} sky-averaged antenna patterns $\mathcal{R}(f)$ for aLIGO (light blue), folded GEO600 (purple), a 600m Michelson interferometer (brown), and a 1200m Michelson interferometer (gray). The folding factor given in Eq.~\eqref{eq:foldedantenna} relates the angular response of an unfolded Michelson interferometer (brown) to its folded counterpart (purple); one can see that the folding factor sharply reduces the angular response at frequencies where it was previously strong. \textbf{Right:} antenna patterns $F(\theta_{\hat n},\phi_{\hat n})$ of folded GEO600 for the following incident-angled GWs: overhead (light blue), completely along one detector arm (gray), and two in-between angles (purple and brown). Here, $F(\theta_{\hat n},\phi_{\hat n}) = \sqrt{|F_+(f;\theta_{\hat n},\phi_{\hat n})|^2 + |F_\times(f;\theta_{\hat n},\phi_{\hat n})|^2}$ since we assume an inclination angle of zero. One can see that the angular behavior of antenna patterns is very frequency dependent at high frequencies. \href{https://github.com/Chris19j/GEO600-High-Frequency-Modeling/blob/main/FIG7-folded-antenna-pattern.ipynb}{\faFileCodeO}
    }
    \label{Fig.7}
\end{figure*}

\subsection{Sky Averaging} \label{subsec:Sky-Averaging}

Let us now calculate the response of the detector to a signal where we average over all possible directions and inclinations. If we define $\Tilde{h}_+(f)=A_+ \Tilde{h}_0(f)$ and $\Tilde{h}_\times(f) = A_\times e^{i \pi/2} \Tilde{h}_0(f)$, the signal measured in the detector would be defined as
\begin{equation}
    \Tilde{h}(f) = Q(f;\theta_{\hat{n}},\phi_{\hat{n}},\iota,\psi) \Tilde{h}_0(f),
\end{equation}
where the angle effects are contained in
\begin{equation}
    \begin{aligned}
        Q(f;\theta_{\hat n},\phi_{\hat n},\iota,\psi) &= F_+(f;\theta_{\hat n},\phi_{\hat n};\psi)A_+ (\iota)
        \\
        &+ i F_\times(f;\theta_{\hat n},\phi_{\hat n};\psi)A_\times (\iota),
    \end{aligned}
\end{equation}
where $A_+ = \frac{1 + \cos^2\iota}{2}$ and $A_\times = \cos\iota$. We care about the absolute value of this squared. We first note that the average of the square of $Q$ over $\psi$ is 
\begin{equation}
    \langle|Q|^2\rangle_\psi = \frac{1}{2}\left(A_+^2 + A_\times^2 \right) \left(|F_+|^2 + |F_\times|^2\right),
\end{equation}
where $F_+ = F_+(\theta_{\hat n},\phi_{\hat n})$ only at $\psi=0$. 
If we average over the direction that the GW arrives, we can write this as 
\begin{equation} \label{A27}
    \langle |Q|^2 \rangle_{\hat n} = \frac{g(\iota)}{2} \mathcal{R}^2(f),
\end{equation}
where $g(\iota) = A_+^2 + A_\times^2$. Now note that 
\begin{align}
    \langle A_+^2 + A_\times^2 \rangle_{\iota} &=  \langle g(\iota)\rangle_{\iota},\\
    &= \frac{1}{4\pi} \int d\Omega_L\left(A_+^2(\iota) + A_\times^2(\iota) \right),\\
    &= \frac{4}{5},
\end{align}
where $d\Omega_L = \sin\iota d\iota d\phi_L$. Now the response function is defined via
\begin{equation} \label{A31}
    \left\langle|F_+|^2 + |F_\times|^2\right\rangle_{\theta_{\hat n},\phi_{\hat n}} = \mathcal{R}^2(f)\,,
\end{equation}
where the DC limit would be given as $\mathcal{R}^2(f) =\frac{2}{5} + \mathcal{O}(\frac{f L}{c})$. This is the sky-averaged antenna response patterns for the plus and cross polarization. Taking the square root of Eq.~\eqref{A31} obtains the sky-averaged antenna patterns used in our SNR calculations:
\begin{equation}\label{C27}
    \mathcal{R}(f) = \sqrt{\left\langle|F_+|^2 + |F_\times|^2\right\rangle_{\theta_{\hat n},\phi_{\hat n}}} \,.
\end{equation}

In Fig.~\ref{Fig.7}, $\mathcal{R}(f)$ is plotted for aLIGO and GEO600. We also compare the response to a Michelson detector of length 600m and 1200m to help clarify how folded detectors differ from normal Michelson detectors. $\mathcal{R}(f)$ is the same shape for aLIGO and the Michelson interferometers, but it is shifted in frequency according to the detector arm length. The folded GEO600 antenna pattern shows the accurate troughs of minimized sensitivity to GWs at odd integer multiples of the FSR, as derived in Eq.~\eqref{eq:nullfoldedfreq}. One can see that the angle-averaged response of folded detectors differs significantly from that of Michelson ones.

\section{Noise Spectral Densities} \label{Appendix-B}

In this appendix, we provide the analytical expressions for the various noise spectral densities that source important contributions to the GEO600 sensitivity function.

\subsection{\label{subsec:quantum-noise}Quantum Noise}
Quantum noise is the combination of quantum shot noise and radiation-pressure noise. Radiation pressure is the optomechanical momentum that high-energy photons give to test masses when they hit them. It is strongest at low frequencies and decreases with frequency at the rate of $\frac{1}{f^2}$ \cite{Schreiber:2018ubt}. Quantum shot noise is derived from the fundamental law that light is discretely quantized, meaning the power is detected from the number of photons $N$ that hit the detector per unit time. Since $N$ is a discrete value, it is calculated using a Poisson distribution, and for large $N$, it becomes a Gaussian with a standard deviation $\sqrt{\langle N \rangle}$. The inherent standard deviation creates a random fluctuation in the number of photons that hit the detector per unit time, creating a fluctuation in the amount of power that is detected \cite{Maggiore:2008ulw}. The small fluctuation in the detected power is called the quantum shot noise, and it is frequency-independent. Our computation of the quantum noise in GEO600 is done directly in Finesse, which has the capability to compute the total quantum noise at some optical readout originating from both photon shot noise and quantum radiation pressure \cite{Freise:2009sf}.

The injection of a squeezed state at the dark port of an interferometer can decrease the noise in one quadrature at the expense of the orthogonal quadrature, which can improve detection sensitivities especially at higher frequencies \cite{carlCaves,Schreiber:2018ubt}.We use the squeezer component built into Finesse to model this procedure; while each distinct model of GEO600 implements frequency-independent squeezing, for each detuning angle of the MSR, we choose the squeezing angle that minimizes the noise at the dark port at the resonant frequency of the SRC.

\subsection{\label{subsec:dark-noise}Dark Noise} 
Dark noise is created by the readout electronics even when no light is hitting the detector. The noise originates from temperature fluctuations inside the ohmic resistance $R$ of the photodiode detection circuit, otherwise known as Johnson noise \cite{Schreiber:2018ubt}.~The ASD of the raw noise is given in dimensions of $\text{W}\!/\!\sqrt{\text{Hz}}$ by \cite{Grote:16} as
\begin{equation}
    A_n^\text{DN} = \sqrt{\frac{4k_BT}{R}},
\end{equation}
where $T$ is the temperature of the resistor. Since dark noise is frequency independent and a factor of $30$ below GEO600's shot noise \cite{Grote:16}, we did not include it in our high-frequency noise budget. 

\subsection{\label{subsec:laser-amplitude-noise}Laser Amplitude Noise}
Laser amplitude noise is power fluctuations created in the laser light during the main laser creation process. We used the relative intensity noise (RIN) ASD, which is given by \cite{Quetschke:2003wia} as
\begin{equation}
    A_{\text{RIN}}(f) = 1.5\cdot10^{-7}\left(\frac{1 \text{Hz}}{f}\right)\frac{1}{\sqrt{\text{Hz}}}.
\end{equation}
Using Eq.~\eqref{eq:strain-noise-PSD} to multiply $A_{\text{RIN}}(f)$ by the necessary transfer functions in Finesse, we obtain the final sky-averaged strain sensitivity for laser amplitude noise.

\subsection{\label{subsec:laser-frequency-noise}Laser Frequency Noise}
Laser frequency noise is created by frequency fluctuations of the main laser light while traveling through the interferometer. The ASD for the raw frequency noise is given in dimensions of $\text{Hz}/\!\sqrt{\text{Hz}}$ by \cite{Kwee:2010fea} as
\begin{equation}
    A_f(f) = \pi f \cdot \sqrt{\frac{2hc}{P\lambda}},
\end{equation}
where $P$ is the laser power hitting the MPR and $\lambda$ is the laser wavelength. Using Eq.~\eqref{eq:strain-noise-PSD} to multiply $A_f(f)$ by the necessary transfer functions in Finesse, we obtain the final sky-averaged strain sensitivity for laser frequency noise.

\subsection{\label{subsec:seismic-noise}Seismic Noise}
Seismic noise is created from external shock waves moving through the ground, causing the mirrors to move. The displacement ASD of the seismic noise in one of GEO600's mirrors without any attenuation is given by \cite{Adams:2014voz} as
\begin{equation}
    A_{x}^\text{SN}(f) = 10^{-7}\left(\frac{1 \text{Hz}}{f^2}\right)\frac{\text{m}}{\sqrt{\text{Hz}}}.
\end{equation}
However, when $f_{gw}>>f_0$, and there are 3 stages of the pendulum, the displacement of the mirror is attenuated by a factor of $({\frac{{f_0}^2}{f^2}})^3$ \cite{Maggiore:2008ulw}. Since GEO600's main mirrors have a fundamental pendulum resonant frequency of $0.5$ Hz \cite{Gossler:2004bbb}, the displacement ASD for the seismic noise in each mirror becomes
\begin{equation}
    A_{x}^\text{SN}(f) = 10^{-7}\left(\frac{1 \text{Hz}}{{f}^2}\right)\left({\frac{{0.5}^2}{f^2}}\right)^3\frac{\text{m}}{\sqrt{\text{Hz}}}.
\end{equation}
Using Eq.~\eqref{eq:strain-noise-PSD} to multiply $A_{x}^\text{SN}(f)$ by the necessary transfer functions in Finesse, we obtain the sky-averaged strain sensitivity of seismic noise for each test mass in GEO600. Adding the seismic strain noise from the BS, MCE, MCN, MFE, and MFN together in quadrature obtains the total sky-averaged strain sensitivity of seismic noise.

\subsection{\label{subsec:thermal-noise}Thermal Noise}
Thermal noise is the most impactful noise source in the middle frequency range $\sim\mathcal{O}(10^2~\text{Hz})$. It affects the central mirrors, far mirrors, and the beam splitter (BS). Thermal noise is calculated separately for the three main parts of each mirror: the substrate, coating, and pendulum suspension. The cause of thermal noise in each of the three parts of the mirrors is due to three different noise sources: Brownian motion, thermo-elastic, and thermo-refractive. For the substrate and coating of the test masses, we calculated the Brownian motion and thermo-elastic noise; the thermo-refractive noise was a factor of $3$ below the thermo-elastic, making it negligible to the phase fluctuation created by the signal $\Tilde{h}(f)$. For the pendulum suspension noise, we only included the fundamental longitudinal eigenmode, exempting any cross-coupling with other pendulum eigenmodes. All pendulum eigenmodes are at negligibly low frequencies relative to our targeted astrophysical sources. For the BS, only the Brownian motion and thermo-refractive noises were calculated - they are the only relevant noise sources created by the BS.

\subsubsection{Test Mass Suspension Thermal Noise}
The suspension thermal noise is the Brownian motion of the molecules moving in the triple pendulum suspending the mirrors. Since the test mass is suspended by a pendulum, the Brownian motion of the suspension has pendulum and violin resonant frequencies where the noise peaks. The displacement ASD for the violin mode noise is given in dimensions of $\text{m}/\!\sqrt{\text{Hz}}$ by \cite{Gossler:2004bbb} as
\begin{equation} \label{suspension-vm}
    A_{x}^\text{VM}(\omega) = \sqrt{\frac{8k_{B}T\omega_0^2\phi_{\text{fiber}~n}}{m\omega[(\omega_n^2 - \omega^2)^2 + \omega_n^4\phi_n^2]}},
\end{equation}
where $\omega=2\pi{f}$, and $\phi_\text{fiber}$ is the loss in the $n$th violin mode:
\begin{equation}
    \phi_{\text{fiber } n} = D_n^{-1} \cdot \left( 1 + \frac{8 d_s}{R} \right) (\phi_{\text{bulk}} + \phi_{\text{nonlin}}).
\end{equation}
$D_n^{-1}$ is the dilution factor given by \cite{Gossler:2004bbb} as
\begin{equation}
    D_n^{-1} = \frac{2}{k_+ L} \left[ 1 + \left( 4 + \frac{(n\pi)^2}{2} \right) \left( \frac{1}{k_+ L} \right) \right],
\end{equation}
where $k_+$ is
\begin{equation}
    k_+ = \sqrt{ \frac{\sqrt{P^2 + 4AEI\rho_L\omega^2 + P}}{2EI}}.
\end{equation}
$P=mg$ is the tension per wire, $\rho_L =\rho \pi R^2$ is the linear mass density of the pendulum fiber, $E$ is Young's modulus of the fiber, and $I=\frac{\pi}{4}R^4$ is the bending moment of inertia of the fiber. $\omega_n$ from Eq.~\eqref{suspension-vm} is the frequency of the $n$th violin-mode oscillation, given by \cite{Gossler:2004bbb} as 
\begin{equation} \label{suspension-eigen-vm}
    \omega_n = \!\frac{n \pi}{L}\! \sqrt{\frac{P}{\rho L}}\! \left[ 1\! + \!\frac{2}{L} \sqrt{\frac{EI}{P}} \! + \!\!\left(\!4 \!+\! \frac{(n\pi)^2}{2} \right) \!\frac{2 EI}{L^2 P} \right].
\end{equation}
Using the parameters from \cite{Gossler:2004bbb}, we obtained the sky-averaged strain sensitivity of the violin mode noise in each mirror by using Eq.~\eqref{eq:strain-noise-PSD}. We calculated the total sky-averaged strain sensitivity of the violin mode noise for GEO600 by adding the noise from every test mass together in quadrature.

The displacement ASD for the pendulum mode noise is given in dimensions of $\text{m}/\!\sqrt{\text{Hz}}$ by \cite{Gossler:2004bbb} as
\begin{equation}
    A_{x}^\text{PM}(\omega) = \sqrt{\frac{4k_{B}T\omega_0^2\phi}{m\omega[(\omega_0^2 - \omega^2)^2 + \omega_0^4\phi^2]}}.
\end{equation}
Using the parameters from \cite{Gossler:2004bbb}, we obtained the sky-averaged strain sensitivity of the pendulum mode noise in each mirror by using Eq.~\eqref{eq:strain-noise-PSD}. Finally, adding the noise from each mirror together in quadrature, we obtained the total sky-averaged strain sensitivity.

\subsubsection{Test Mass Substrate Thermal Noise}
We simulated the Brownian motion and thermo-elastic noise within the substrate of the test mass mirrors. The displacement ASD for the Brownian motion substrate noise is given in dimensions of $\text{m}/\!\sqrt{\text{Hz}}$ by \cite{Heinert:2014pba} as
\begin{equation}
    A_{x}^\text{TMBM}(\omega) = \sqrt{\frac{8k_BT\mathbb{E}\phi}{\omega{F_0}^2}},
\end{equation}
where $\mathbb{E}$ is the mean elastic energy stored in the test mass. For the substrate, the mean elastic energy is given by \cite{Heinert:2014pba} as 
\begin{equation}
    \mathbb{E}_{\text{Substrate}} = \frac{1-\sigma^2}{2\sqrt{2\pi}Yr_0}\cdot{F_0}^2.
\end{equation}
Using the parameters from \cite{Heinert:2014pba}, we obtained the sky-averaged strain sensitivity of the Brownian motion substrate noise in each mirror by using Eq.~\eqref{eq:strain-noise-PSD}. We added the noise from each individual mirror together in quadrature to obtain the total sky-averaged strain sensitivity for the Brownian motion in the substrate of GEO600's mirrors.

For the thermo-elastic noise, the displacement ASD is given in dimensions of $\text{m}/\!\sqrt{\text{Hz}}$ by \cite{Heinert:2014pba} as 
\begin{equation}
    \hspace{-0.5em} 
    A_{x}^\text{TMTE}(\omega) = 
    \sqrt{
        \begin{aligned}
            &\left[
                1+\frac{k_sr_0}{\sqrt{2\pi}}
            \right] \times \\
            &\left[
                \frac{8\kappa\alpha^2\left(1+\sigma\right)^2k_BT_0^2}
                {\sqrt{2\pi}\rho^2C_p^2r_0^3\omega^2} 
            \right]
        \end{aligned}
    }.
\end{equation}
Using the parameters from \cite{Heinert:2014pba}, we obtained the sky-averaged strain sensitivity of the thermo-elastic noise in the substrate of each mirror by using Eq.~\eqref{eq:strain-noise-PSD}. We added the noise from each individual mirror together in quadrature to obtain the total sky-averaged strain sensitivity for the thermo-elastic noise in the substrate of GEO600's mirrors.

\subsubsection{Test Mass Coating Thermal Noise}
The coating thermal noise we simulated was the Brownian motion in each reflective coating layer and the thermo-elastic noise created by temperature fluctuations within the mirror coatings. The noise caused by the Brownian motion in the mirror coatings is given in dimensions of $\text{m}/\!\sqrt{\text{Hz}}$ by \cite{Heinert:2014pba} as
\begin{equation}
    A_{x}^\text{TMBM}(\omega) = \sqrt{\frac{8k_BT\mathbb{E}\phi}{\omega{F_0}^2}},
\end{equation}
where the mean elastic energy $\mathbb{E}$ for the coating is calculated as
\begin{equation}
    \begin{aligned}
        \hspace{-0.5em} 
        \mathbb{E}_{\text{Coating}} = \frac{HF_0^2}{4\pi{r_0^2}}\biggl[\frac{(1+\sigma_c)(1-2\sigma_c)}{Y_c(1-\sigma_c)}\\
        + \frac{Y_c(1+\sigma)^2(1-2\sigma)^2}{Y^2(1-\sigma_c^2)}
        \biggr].
    \end{aligned}
\end{equation}
Using the parameters from \cite{Heinert:2014pba}, we obtained the sky-averaged strain sensitivity of the Brownian motion in the coating of each of GEO600's mirrors by using Eq.~\eqref{eq:strain-noise-PSD}. We added the noise from each individual mirror together in quadrature to obtain the total sky-averaged strain sensitivity for the test-mass coating Brownian motion noise.

\begin{figure*}[t]
    \centering
    \includegraphics[width=0.49\textwidth]{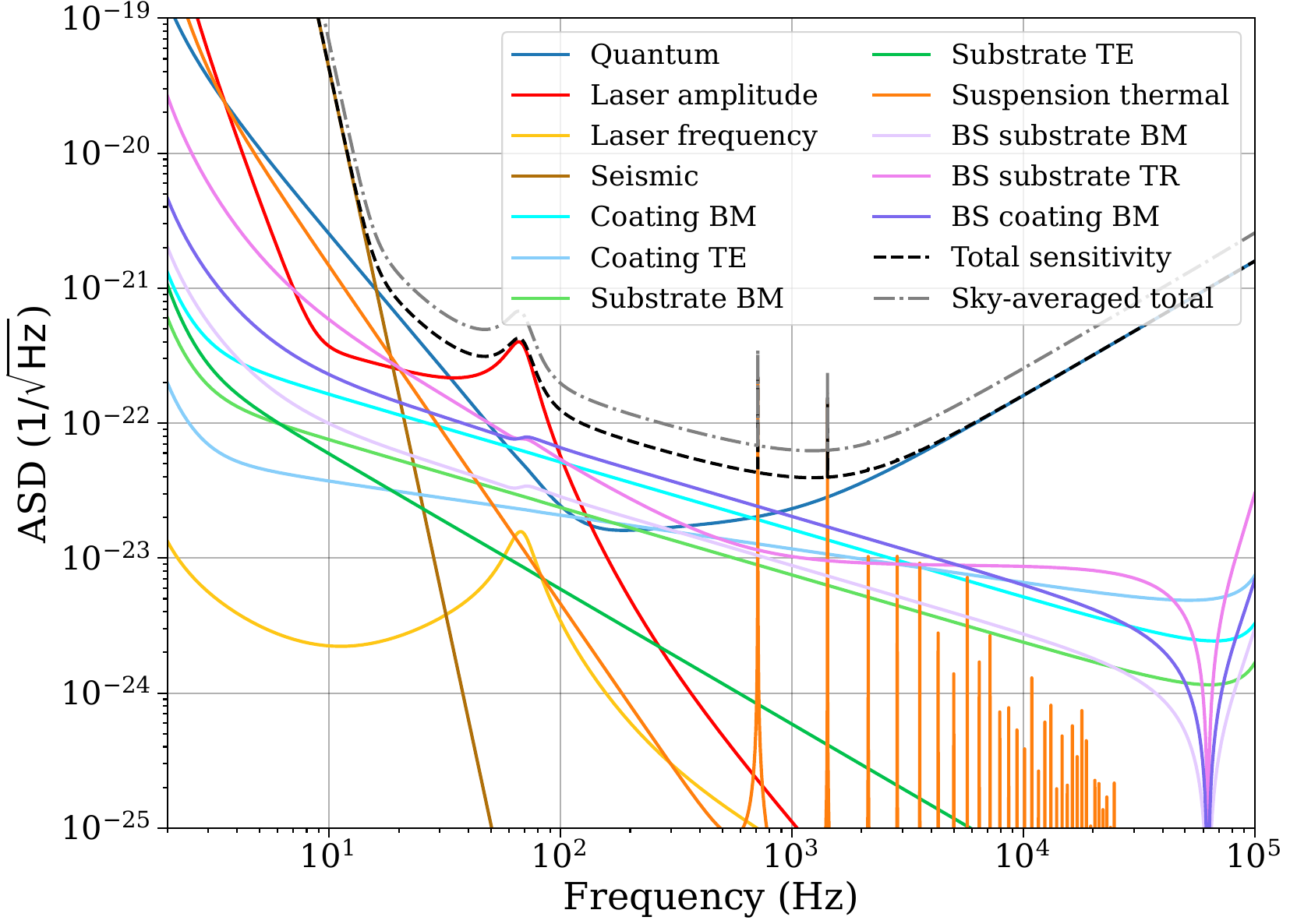}
    \includegraphics[width=0.49\textwidth]{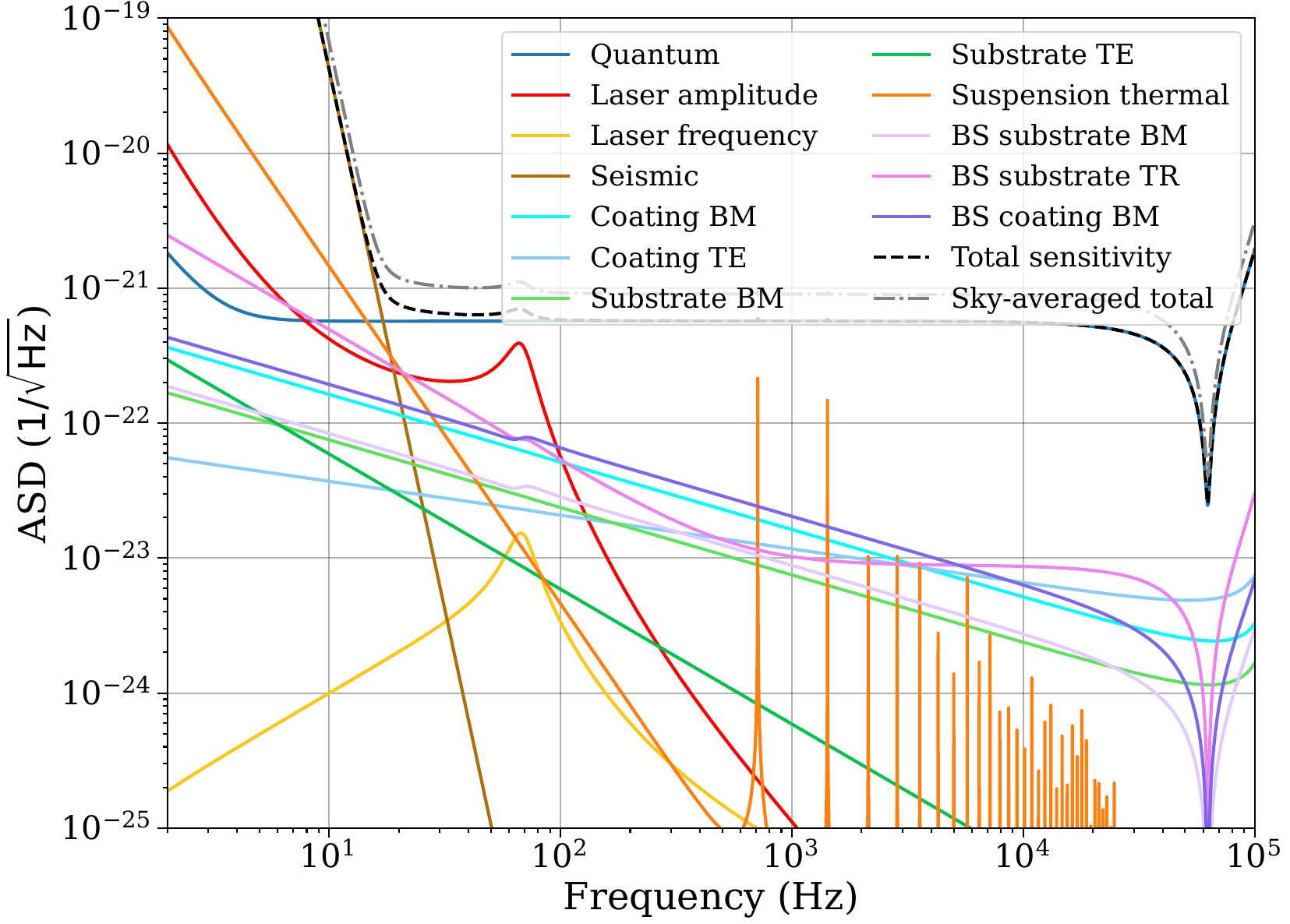}
    \caption{Strain sensitivity $ A_h(f) $ of GEO600 with a tuned MSR (left) and an anti-tuned MSR (right), including all modeled noise sources and the total sensitivity, both with and without sky-averaging. Note that all individual sources, as well as the total sensitivity without sky-averaging, assume a normal-incidence GW sky location. BM: Brownian motion, TE: thermo-elastic, TR: thermo-refractive. \href{https://github.com/Chris19j/GEO600-High-Frequency-Modeling/blob/main/FIG8-GEO-noise-budget.ipynb}{\faFileCodeO}
    }
    \label{Fig.8}
\end{figure*}

The displacement ASD of the thermo-elastic coating noise is given in dimensions of $\text{m}/\!\sqrt{\text{Hz}}$ by \cite{Heinert:2014pba} as
\begin{equation}
    A_{x}^\text{TMTE}(\omega) = \sqrt{\frac{3\sqrt{2}\alpha^2Y_c^2M^2k_BT_0^2H^2}{2\pi\sqrt{\kappa\rho{C_p}}r_0^2\sqrt{\omega}}}.
\end{equation}
Using the parameters from \cite{Heinert:2014pba}, we obtained the sky-averaged strain sensitivity of the coating thermo-elastic noise in each mirror by using Eq.~\eqref{eq:strain-noise-PSD}. We added the noise from each individual mirror together in quadrature to obtain the total sky-averaged strain sensitivity for the thermo-elastic noise in GEO600's test mass coatings.
 
\subsubsection{Beam Splitter Thermal Noise}

The absence of Fabry-Perot cavities in GEO600 allows kilowatts of power to pass through the BS as light propagates in the PRC, creating the most dominant noise sources in the hundreds of Hz. The high levels of power create pockets of fluctuating temperature within the BS's substrate, changing the index of refraction and adding extra phase to the light \cite{Benthem:2009fk}.~In addition to the substrate thermo-refractive noise, we calculated the coating Brownian motion and substrate Brownian motion noises within the BS \cite{Dickmann:2018wng}. The displacement ASD of the thermo-refractive noise in the BS's substrate is given in dimensions of $\text{m}/\!\sqrt{\text{Hz}}$ by \cite{Benthem:2009fk} as 
\begin{equation}
    \hspace{-0.5em}
    A_x^\text{BSTR}(\omega) = \!\!\!
    \sqrt{
        \begin{aligned}
            &\!\!\left[
                \frac{4{k_B}{\kappa}{T^2}{\beta^2}a'\left(\eta+\eta^{-1}\right)}{\pi\left(C{\rho}{r_0}^2\omega\right)^2 2\eta^2}
            \right] \times \\
            &\!\!\left[
                1 + \frac{2{k^2}{r_0}^2\eta}{\left(\eta+\eta^{-1}\right)\left(1\!+\!\left(2kl_{th}(\omega)\right)^4\right)}
            \right].
        \end{aligned}
    }
\end{equation}
Using the parameter values that represent GEO600 from \cite{Benthem:2009fk}, we obtained the sky-averaged strain sensitivity of the thermo-refractive noise in the BS's substrate by using Eq.~\eqref{eq:strain-noise-PSD}.

The displacement ASD of the Brownian motion in the BS is given in dimensions of $\text{m}/\!\sqrt{\text{Hz}}$ by \cite{Dickmann:2018wng} as
\begin{equation}
    A_{x}^\text{BSBM}(\omega) = \sqrt{\frac{8k_BT\mathbb{E}\phi}{\omega{F_0}^2}}.
\end{equation}
For the substrate, the mean elastic energy is given by \cite{Dickmann:2018wng} as
\begin{equation}
    \mathbb{E}_{\text{Substrate}} = \left(1.97 \cdot 10^{-9}\right)\cdot{F_0}^2.
\end{equation}
Plugging both $\mathbb{E}$ and the parameter values from \cite{Dickmann:2018wng} into $A_{x}^\text{BSBM}(\omega)$—and then using Eq.~\eqref{eq:strain-noise-PSD}—we obtained the sky-averaged strain sensitivity for the Brownian motion noise within the BS's substrate. 

For the BS coating, we found the mean elastic energy by solving for $\mathbb{E}$ given the stated outcome of the PSD at $100$ Hz given by \cite{Dickmann:2018wng}:
\begin{equation}
    \mathbb{E}_{\text{Coating}} = \left(4.91 \cdot 10^{-13}\right)\cdot{F_0}^2.
\end{equation}
Plugging both $\mathbb{E}$ and the parameter values from \cite{Dickmann:2018wng} into $A_{x}^\text{BSBM}(\omega)$—and then using Eq.~\eqref{eq:strain-noise-PSD}—we obtained the sky-averaged strain sensitivity for the Brownian motion noise within the coating of the BS. The sky-averaged strain sensitivity $ \langle A_h(f) \rangle $ of every noise sources we modeled, as well as the total sky-averaged strain sensitivity of the interferometer, are shown in an ASD plot in Fig.~\ref{Fig.8}.

\newpage

\bibliography{References.bib}

\end{document}